\documentclass[review]{elsarticle}

\usepackage[utf8]{inputenc}
\usepackage{lineno,hyperref}
\modulolinenumbers[5]
\usepackage{amsmath}
\usepackage{amssymb}
\usepackage{amsfonts}
\usepackage{graphicx}
\numberwithin{equation}{section}

\journal{Journal of \LaTeX\ Templates}

\begin{document}

\begin{frontmatter}

\title{Monopole Black Hole in Stringy gravity}

\author{Hongsu Kim\fnref{myfootnote}}
\address{Center for Theoretical Astronomy, Korea Astronomy and Space Science Institute, Daejeon 34055, Republic of Korea}
\fntext[myfootnote]{chris@kasi.re.kr}

\begin{abstract}
The possible transition of large mass magnetic monopole solutions to monopole black hole solutions in Yang-Mills-Higgs theory with spontaneous symmetry breaking (SSB) coupled to the low energy effective theory of superstring with only dilatonic sector is studied. Our main motivation is to explore the effects of the dilaton field on the monopole black hole solutions. Working in the Einstein conformal gauge, it is explicitly shown, in terms of the Hawking evaporation of the monopole black hole, that the presence of the dilaton field appears to introduce a "mass dependent extra attractive interaction" in addition to that of the standard Einstein gravity into the system.
\end{abstract}

%\begin{keyword}

%\end{keyword}

\end{frontmatter}

 \section{introduction}
 The interest in non-perturbative soliton solutions of classical nonlinear field theories coupled to gravity can be traced back as early as to min-seventies. For instance, 't Hooft magnetic monopole solution in Yang-Mills-Higgs theory with spontaneous symmetry breaking (SSB) coupled to Einstein gravity had been studied by several authors \cite{1,2}. It had been found that magnetic monopole solution of 't Hooft-Polyakov type \cite{4} in curved spacetime exists with the exterior spacetime being represented by the Reissner-Nordstr{\"o}m metric (corresponding to a magnetic charge $Q=1/e$) for "small mass" magnetic monopoles.
 
 Recently, however, this curved spacetime magnetic monopole have received revived attention due to their possible transition to black hole solutions for "large mass" magnetic monopoles, namely "monopole black holes". \cite{3,5}
 
 Intuitively, the possibility of transition to the monopole black holes can be understood as follows; as the Higgs field vacuum expectation value $v$ is "varied", the mass of the monopole $M_{mon} \sim v/e$ and the size of the monopole, $R_{mon} \sim1/ev$ are also varied (where $e$ denotes the gauge coupling constant related to the magnetic charge by $Q=1/e$ as mentioned earlier.)
 Then for large enough value of the Higgs field VEV, $ v$ (such as $v \ge M_{pl}$), it would be possible to have $2GM_{mon}/R_{mon} \sim (\frac{v}{M_{pl}})^2 \ge 1$ implying that the corresponding monopole solution should be a "black hole" since the Schwarzschild radius $2GM_{mon}$ becomes comparable to the monopole radius $R_{mon}$.
 
 On the other hand, there also has been a number of studies on the classical solutions to the low energy effective theory of superstring such as black hole solutions \cite{5} and soliton solutions \cite{6}, lately. Particularly, it has been pointed out \cite{7} that the $N=4$ supersymmetric low energy effective theory of superstring compactified down to (3+1) dimensions allows the magnetic monopole solution of 'BPS' type (namely, Bogomol'nyi-Prasad-Sommerfield solution) with the metric solution being well-behaved everywhere, i.e. having no singularities whatsoever. However, since the non-supersymmetric low energy theories of superstring are known to allow singular metric solutions, i.e. charged black hole solutions \cite{5}, it appears that extremal supersymmetric solutions (classical) to low energy effective theory of superstring are better behaved than non-supersymmetric solutions, in general. In the present work, we would like to study the "large mass" magnetic monopole solutions in the Yang-Mills-Higgs theory with SSB coupled to the (3+1) dimensional low energy effective theory of bosonic part of superstring \cite{8} with all the bosonic degrees of freedom except for the metric $g_{\mu\nu}$ and the dilaton field $\Phi$ being set to zero. A theory like this might not look so compelling but our motivation is to compare the properties of classical solutions (in particular, black hole solution) of our theory with those of the Einstein-Yang-Mills-Higgs system \cite{3} especially in order to explore the effects of the dilaton field on the properties of classical solutions. (In fact, in a sense, the Einstein-Yang-Mills-Higgs theory is less compelling than our theory is since at very small length scales where matter fields have non-trivial quantum behaviors, Einstein gravity reveals bad short-distance behavior (e.g. ultraviolet divergences, 'the graceful exit' problem in old inflation model and 'the large wormhole' problem).)
 We find that the presence of dilaton field requires larger critical mass for the transition to "monopole black hole" to occur and that it also speeds up the termination of the black hole evaporation via Hawking radiation process when compared with the Einstein gravity case. Therefore, it appears that the presence of dilaton field turns out to introduce a mass-dependent extra attractive interaction as we shall see.
 
 Finally, we end by pointing out an interesting possible cosmological implication of monopole black holes. One of the currently puzzling and unsettled issues in cosmology is the identification of "cold dark matter" associated with the missing mass problem of the universe. (According to the inflationary universe scenario, for instance, the mass density of the universe today should be almost equal to the "critical mass" $\rho_c=\frac{3H^2_o}{8\pi G}$ (where $H_o$ is the present value of the Hubble parameter) for spatially-flat universe. This statement, when compared with the present cosmological observations, lead us to conclude that about 90 to 99 percent of the universe mass density is "missing" which necessarily demands the existence of "dark matter".) If the transition of "large mass" magnetic monopoles to monopole black holes is indeed quite possible, then these non-Abelian monopole black holes could be a good candidate for a form of cold dark matter and also it is no surprise that non-Abelian magnetic monopoles predicted to exist in unified particle theories \cite{9} have never been seen thus far.
 
 \section{Formulation of the theory}
 As mentioned earlier, we are interested in the magnetic monopole solutions in the Yang-Mills-Higgs theory with SSB coupled to the low energy effective theory of superstring.
Therefore, for the gravity sector of our theory, we take the gravity action obtained from the low energy effective theory of bosonic part of superstring compactified to 4-dimensions \cite{8}
 \begin{equation}
 S_G=\int d^4x\sqrt{g} e^{-2\Phi(x)}\left[\frac{1}{2k^2}R-\Lambda+\frac{2}{k^2}(\nabla\Phi)^2\right] 
 \end{equation}
 where $k^2\equiv8\pi G$, $\Lambda$ is the cosmological constant which will be set to zero in the actual calculations later and $\Phi(x)$ denotes the "dilaton" field. Here, also note that we have set the remaining rank-two gauge field $F_{\mu\nu}$ and rank-three antisymmetric tensor field $H_{\mu\nu\lambda}$ $(H=dB-w^o_{3Y}+w^o_{3L}$ where $w$'s are gauge and Lorentz Chern-Simons three-forms) to zero in order to explore the effects of dilaton field alone.
 
 And for the matter sector, we take the familiar Yang-Mills-Higgs theory with SSB which, in flat spacetime, is known to admit the magnetic monopole solution of 't Hooft-Polakov type \cite{4}
 \begin{equation}
 S_M=\int d^4x\sqrt{g}\left[-\frac{1}{4}(F^a_{\mu\nu})^2-\frac{1}{2}(D_\mu\phi^a)^2-U(\phi^a\phi^a)\right]
 \end{equation}
 where
 \begin{align*}
 F^a_{\mu\nu}&=\partial_\mu A^a_\nu-\partial_\nu A^a_\mu+e\epsilon^{abc}A^b_\mu A^c_\nu,\\
 D_\mu\phi^a&=(\partial_\mu\delta^{ac}+e\epsilon^{abc}A^b_\mu)\phi^c,\\
 U(\phi^a\phi^a)&=\frac{\lambda}{4}(\phi^a\phi^a-v^2)^2 \hspace{5mm} 
 \end{align*}
 and $a, b, c=1, 2, 3.$
 
 Note that we have added a "constant vacuum energy" term $\frac{\lambda}{4}v^4$ to the Higgs field potential so that the energy vanishes in the broken-symmetry vacuum. Latin indices a, b, c refer to the internal $SU(2)$ gauge group indices and as a consequence of SSB, a neutral Higgs field turns out to have the mass $m_H=\sqrt{\lambda}v.$\\
 Now putting the gravity and matter sector together, our theory is represented by the total action,
 \begin{equation}
 S=\int d^4x\sqrt{g}\left\{e^{-2\Phi(x)}\frac{1}{2k^2}[R+4(\nabla\Phi)^2]-\frac{1}{4}(F^a_{\mu\nu})^2-\frac{1}{2}(D_\mu\phi^a)^2-U(\phi^a\phi^a)\right\}
\end{equation}
consider the "Weyl rescaling" of gravitational fields (i.e. metric and dilaton field)
\begin{equation}
\begin{split}
&g_{\mu\nu}=e^{2\Phi(x)}\tilde{g}_{\mu\nu}(x),\\
&\Phi(x)=\tilde{\Phi}(x).
\end{split}
\end{equation}
(the conformal weight of the dilaton field $\Phi$ is zero since $\Phi(x)$ is "dimensionless" in the gravity action written in "sigma model" form.)\\
Then under the Weyl rescaling the action transforms into the "Einstein conformal gauge" form,
 \begin{equation}
 \begin{split}
 S=\int d^4x\sqrt{g}&\left\{\frac{1}{16\pi G}[R-2(\nabla\Phi)^2]\right.\\
 &\left.-\frac{1}{4}(F^a_{\mu\nu})^2-\frac{1}{2}e^{2\Phi(x)}(D_\mu\phi^a)^2-e^{4\Phi(x)}U(\phi^a\phi^a) \right\}
 \end{split}
 \end{equation}
 where we dropped tilde.
Note, first, that the dilaton field $\Phi(x)$, which used to be a ghost field with negative definite norm, now takes the kinetic term of the canonical form and hence turns into a physical scalar field (with positive definite norm) in this Einstein conformal gauge after the Weyl rescaling of metric and dilaton field.

 We will work in this "Einstein conformal gauge" in which the Einstein-Hilbert action rather than some multiple of it by a non-linear realization of the dilaton field appears. We choose this conformal gauge because physical interpretations such as the behavior of horizons and black hole thermodynamics are more conveniently discussed in this gauge.
 
 Now by varying the action with respect to the Higgs field $\phi$, the Yang-Mills field $A^a_\mu$, the dilaton field $\Phi$ and the metric $g_{\mu\nu}$ respectively, we obtain the following field equations
 \begin{equation}
 \begin{split}
 &\frac{1}{e^{2\Phi}\sqrt{g}}D_\mu\left[e^{2\Phi}\sqrt{g}g^{\mu\nu}(D_\nu\phi)^a\right]-e^{2\Phi}\frac{\partial U(\phi^b\phi^b)}{\partial \phi^a}=0,\\
 &\frac{1}{\sqrt{g}}D_\mu[\sqrt{g}F^{a\mu\nu}]-e^{2\Phi}[e\epsilon^{abc}\phi^b(D^\nu\phi^c)]=0,\\
 &\square\Phi-4\pi Ge^{2\Phi}[g^{\mu\nu}(D_\mu\phi)^a(D_\nu\phi)^a+4e^{2\Phi}U(\phi^b\phi^b)]=0,\\
 &R_{\mu\nu}=2\partial_\mu\Phi\partial_\nu\Phi\\&\hspace{1cm}+8\pi G\left[\{g^{\rho\sigma}F^a_{\mu\rho}F^a_{\nu\sigma}-g_{\mu\nu}(\frac{1}{4}g^{\alpha\rho}{g^{\beta\sigma}F^a_{\rho\sigma}F^a_{\alpha\beta})\}}\right.\\
&\hspace{1cm}\left.+e^{2\Phi}\{(D_\mu\phi)^a(D_\nu\phi)^a+g_{\mu\nu}e^{2\Phi}U(\phi^b\phi^b)\}\right]
 \end{split}
 \end{equation}
 where $D_\mu$ denotes gauge covariant derivative as defined earlier and $\square\equiv\frac{1}{\sqrt{g}}\partial_\mu[\sqrt{g}g^{\mu\nu}\partial_\nu].$
 
 \section{Ansatz for the solutions}
 \subsection{Ansatz for matter field solutions}
 we look for static, spherically symmetric solutions to these field equations that are asymptotically flat. Thus the metric can be written in the form
 \begin{equation}
 ds^2=-B(r)dt^2+A(r)dr^2+r^2d\Omega^2_2
 \end{equation}
where $d\Omega^2_2$ is the metric on the unit two sphere and B($\infty$)=1 for the normalization of t and A($\infty$)=1 due to the asymptotic flatness condition.

For the matter sector, in order to look for a spherically symmetric topological soliton solution, i.e. magnetic monopole solution in our theory, we take, as a starting point, the standard ansatz for scalar field solution and gauge field solution which are the same in form as the flat spacetime 't Hooft-Polyakov monopole solution ansatz. Namely, we begin by assuming that the soliton solution is spherically symmetric in order to greatly simplify the task of finding an explicit solution. In a gauge theory, especially in Yang-Mills theory with the gauge group $SU(2) \sim SO(3)$, however, it is not sensible to demand more than spherical symmetry up to a gauge transformation. For example, the scalar field configuration $\phi^a(x)$ is said to be spherically symmetric if the effect of a spatial rotation of $\phi^a(x)$ can be compensated by a gauge transformation. Thus from the asymptotic flatness condition, we expect as in the case of flat spacetime that the asymptotic behavior of $\phi^a(x)$ and $A^a_i(x)$ is invariant under a simultaneous spatial rotation and global $SU(2)$ gauge transformation. We also assume that this invariance and the parity invariance
\[x^i \rightarrow -x^i, \quad \phi^a \rightarrow -\phi^a, \quad A^a_i \rightarrow -A^a_i\]
hold for all $x$. Then we arrive at the "standard" ansatz \cite{4} in terms of Cartesian coordinates
\begin{align}
\begin{split}
\phi^a(x)&=v\frac{x^a}{r}h(r),\\ 
A^a_0(x)&=0,\\
A^a_i(x)&=-\epsilon_{iab}\frac{x^b}{er^2}[1-u(r)] 
\end{split}
\end{align}
with the boundary conditions $h(0)=0,\ u(0)=1$ and $h(\infty)=1,\ u(\infty)=0$ for the non-singular monopole solution at the origin and the asymptotic flatness (or equivalently the finite energy solutions) condition respectively.

In the following, then, we shall see that this ansatz for matter fields indeed provides an exact magnetic monopole solution which satisfies the above boundary condition. To this end, we first consider the classical field equations for the Higgs field and for the Yang-Mills field in terms of the spherically-symmetric ansatz in spherical-polar coordinates:
\begin{align}
\begin{split}
&\frac{1}{r^2\sqrt{AB}e^{2\Phi(r)}}\left[\frac{r^2\sqrt{AB}h'(\nu)}{A}e^{2\Phi(r)}\right]'-\frac{2}{r^2}hu^2-\lambda v^2h(h^2-1)e^{2\Phi(r)}=0,\\
&\frac{1}{\sqrt{AB}}\left[\frac{\sqrt{AB}u'}{A}\right]'-\frac{u(u^2-1)}{r^2}-e^2v^2uh^2e^{2\Phi(r)}=0.
\end{split}
\end{align} 
Note that an exact, albeit singular, solution of these matter field equations exists "independently" of the gravitational fields (i.e. metric fields $B(r), A(r)$ and it is
\begin{equation}
h(r)=1, \  u(r)=0
\end{equation}
which corresponds to the asymptotic (as $r\rightarrow\infty$) form of the 't Hooft'-Polyakov magnetic monopole solution \cite{4}. Also note that this asymptotic solution of the matter field equations actually satisfies the field equations everywhere like in the case of flat spacetime as was first observed by Bais and Russell, and by Cho and Freund \cite{1}. What is new here is that the magnetic monopole solution of 't Hooft-Polyakov type still exists even in the presence of the dilaton field arising in the low energy effective theory of superstring.

\subsection{Ansatz for the dilaton field as such}
In this time, in order to find the dilaton solution, we substitute the ansatz for the Higgs field solution and the Yang-Mills field solution (which, as we have seen in the previous section, actually yields an exact monopole solution) into the classical field equation for the dilaton. And by using the asymptotic behavior (as $r\rightarrow\infty$) of this monopole solution and the asymptotic flatness of the metric, one would be able to determine the asymptotic behavior of the dilaton solution and further the ansatz for the dilaton field that is valid for all $r$. Thus to this end, we consider the dilaton field equation in terms of the monopole solution ansatz in spherical-polar coordinates

\begin{subequations}
\begin{align}
&\frac{1}{r^2\sqrt{AB}}\left[\frac{r^2\sqrt{AB}\Phi'(r)}{A}\right]'\\
&-4\pi Ge^{2\Phi(r)}\left\{\left[\frac{1}{	A}v^2(h')^2+2\frac{v^2h^2u^2}{r^2}\right]+4e^{2\Phi(r)}\left[\frac{\lambda}{4}v^4(h^2-1)^2\right]\right\}=0.\nonumber\\
&\text{As mentioned, we begin by substituting the exact monopole solution into}\nonumber\\ 
&\text{this dilaton field equation, i.e. inserting $h(r)=1, u(r)=0$ yields}\nonumber \\
&\frac{1}{r^2\sqrt{AB}}\left[\frac{r^2\sqrt{AB}}{A}\Phi'(r)\right]'=0.
\end{align}
\end{subequations}
Since the exact monopole solution found actually represent asymptotic form (as r $\rightarrow \infty$) of the monopole solution, one would obtain the asymptotic dilaton solution by further putting in the "asymptotic flatness" condition namely, $B(r) \rightarrow 1,\  A(r) \rightarrow 1$ as $r \rightarrow \infty$. Now one has $\frac{1}{r^2}\frac{\partial}{\partial r}\left[r^2\frac{\partial\Phi}{\partial r}\right] \simeq 0$ as $r \rightarrow \infty$. Namely for large r, the dilaton field satisfies the "source-free" Poisson equation. Then by letting $\Phi(r)\equiv \frac{\psi(r)}{r}$ the above equation reduces to $\frac{d^2\psi}{dr^2}=0$ whose general solution is $\psi(r)=a+br$ (where a, b are integration constants). Thus we have $\Phi(r)=\frac{a}{r}+\Phi_\infty.$
Note here that firstly we have set $b=\Phi_\infty$ since the integration constant $b$ should represent the "asymptotic constant" value of the dilaton field, $\Phi_\infty.$ Secondly, since the dilaton field appears in the gravity action in non-linear sigma model form, it is dimensionless and hence the other integration constant $a$ should have the mass dimension of -1.

Here, we choose the arbitrary mass parameter associated with this integration constant $a$ to be the same as the arbitrary mass parameter that would arise in solving the Einstein field equations as we shall see in the next section. This is because all the field equations (matter and gravitational) are coupled and hence, in principle, should be solved simultaneously with the consistent choice of integration constants and because obviously the theory should involve only one arbitrary mass parameter, i.e. the mass of the monopole solution, M. Therefore, we take $a=\frac{1}{M}$ (note also that since there is no coupling between the Yang-Mills and the dilaton field in the action, the field equation for the dilaton has the source term which has no dependence on the Yang-Mills field. Thus the dilaton solution should not have any dependence on the magnetic charge $\frac{1}{e}$). Then the asymptotic dilaton solution is given by $\Phi(r)=\frac{1}{Mr}+\Phi_\infty$ \ as \ $r \rightarrow \infty.$
Further, one may wish to construct an ansatz for the dilaton field based on this asymptotic behavior. Namely, one may well take the ansatz for the dilaton solution as 
\begin{equation}
\Phi(r)=\frac{1}{Mr}w(r)+\Phi_\infty
\end{equation}
with the boundary conditions $w(r) \rightarrow (\Phi_0-\Phi_\infty)Mr$ as $r\rightarrow 0$ (for non-singular dilaton solution) and $w(r)\rightarrow1$ as $r\rightarrow \infty$. And here $\Phi_0$ is some constant which represents the correct value of the dilaton field near the origin. Now we arrive at the "self-consistent" ansatz for all the fields present in our theory.

Finally we mention that in our theory, the additional gravitational degree of freedom, i.e. the dilaton field does not give rise to any new parameter associated with it (namely, the "dilaton charge") other than the mass parameter $M$ and the magnetic charge $\frac{1}{e}$ to characterize classical solutions of our theory.
In other words, at spatial infinity, the dilaton field in our theory is given by $\Phi(r)=\frac{1}{Mr}+\Phi_\infty$. This asymptotic behavior then allows us to compute the dilaton charge defined by the Gauss' law as:
\begin{equation}
D\equiv\frac{1}{4\pi}\int d\Sigma^\mu\nabla_\mu\Phi=-\frac{1}{M}.
\end{equation}
Therefore, obviously the dilaton charge is not a new, free parameter since it is determined by the mass parameter. Note. however, that the dilaton charge is always "negative" and also that it is determined not by the magnetic charge but only by the mass parameter because there is no direct coupling between the dilaton and the Yang-Mills field whereas there is the coupling between the dilaton and the Higgs field in the action as stated earlier. Here, particularly note that the dilaton charge is negative and inversely proportional to $M$. Before closing this section, since we have constructed a self-consistent ansatz for all the fields present, we write the "non-metric" sector of the action in terms of this ansatz in spherical-polar coordinates
\begin{equation}
S_{\text{non-metric}}=-4\pi\int dt\ dr\ r^2\sqrt{AB}\left[\frac{1}{A}K(w,u,h)+V(w,u,h)\right]
\end{equation}
where
\begin{align*}
K(w,u,h)&\equiv\frac{1}{8\pi GM^2}(\frac{w'}{r}-\frac{w}{r^2})^2+\frac{(u')^2}{e^2r^2}+e^{2(\frac{1}{Mr}w+\Phi_\infty)}\frac{1}{2}v^2(h')^2\\
V(w,u,h)&\equiv\frac{(u^2-1)^2}{2e^2r^4}+e^{2(\frac{1}{Mr}w+\Phi_\infty)}\frac{v^2h^2u^2}{r^2}+e^{4(\frac{1}{Mr}w+\Phi_\infty)}
\frac{\lambda}{4}v^4(h^2-1)^2.
\end{align*}
Note that $K$ and $V$ are positive-definite and do not have explicit dependence on the metric.

\section{Solution to Einstein equations}
Since we have the self-consistent, static spherically-symmetric ansatz for fields, we now substitute it into the Einstein equations and attempt to solve them.
Only two components of the Einstein field equations out of the three are truly independent because the third component is satisfied automatically due to the energy-momentum conservation $T^{\mu\nu}_{;\mu}=0$. Thus we consider the following two independent combinations convenient in solving the Einstein equations;
\begin{equation}
\begin{split}
&\frac{1}{AB}(AR_{tt}+BR_{rr})=8\pi G[-T^t_t+T^r_r],\\
&\frac{1}{2}\left(\frac{1}{B}R_{tt}+\frac{1}{A}R_{rr}\right)+\frac{1}{r^2}R_{\theta\theta}=8\pi G[-T^t_t].
\end{split}
\end{equation} 
In terms of the ansatz, they become
\begin{equation}
\begin{split}
&\frac{(AB)'}{AB}=16\pi GrK(w,u,h),\\
&A(r)=\frac{1}{\left[1-\frac{2GM(r)}{r}\right]}\\
\end{split}
\end{equation} 
with $M'(r)\equiv4\pi r^2\left[\frac{1}{A}K(w,u,h)+V(w,u,h)\right].$\\

Now, we would like to find the metric solution that describes the exterior spacetime of the monopole configuration. To do so we substitute the asymptotic behaviors (as r $\rightarrow\infty$) of the solutions of non-metric fields (dilaton, Higgs and Yang-Mills field) into the Einstein equations above, i.e. insert $w(r)\rightarrow1$, $h(r)\rightarrow1$ and $u(r)\rightarrow0$ as $r\rightarrow\infty$.\\
Then from the first combination, we have 
\[
\frac{(AB)'}{AB}=2\frac{1}{M^2r^3}
\]
 which is readily integrated to yield
\begin{equation}
B(r)=\exp\left[-\frac{1}{M^2r^2}\right]A^{-1}(r)
\end{equation}
where we set the irrelevant integration constant to zero. Next, from the second combination, we have
\begin{equation}
M'(r)+\frac{1}{M^2r^3}M(r)-4\pi\bigg(\frac{1}{8\pi GM^2}+\frac{1}{2e^2}\bigg)\frac{1}{r^2}=0.
\end{equation}
Unfortunately, this differential equation con not be readily integrated to give a simple form of $M(r)$. However, since we are essentially interested in the "exterior metric solution" at large $r$ we can approximate the metric solution systematically. Namely we may use the "interaction" method that allows us to find the corrections to the metric solution as one moves from spatial infinity inward. Obviously, then, the leading approximation would be the scheme in which one sets $A(r)\simeq1$ in the source form (i.e. the stress tensor term) on the right hand side of Einstein equations. This, in turn, is equivalent to neglecting the second term $\frac{M(r)}{M^2r^3}$ in the differential equation for $M(r)$ above. Thus we have, to leading order,
\begin{equation}
\begin{split}
&M(r)\simeq M-2\pi\bigg(\frac{1}{e^2}+\frac{1}{4\pi GM^2}\bigg)\frac{1}{r}+\mathcal{O}\bigg(\frac{1}{r^2}\bigg),\\
&A(r)\simeq\left[1-\frac{2GM}{r}+\bigg(\frac{4\pi G}{e^2}+\frac{1}{M^2}\bigg)\frac{1}{r^2}\right]^{-1}
\end{split}
\end{equation}
where we chose the integration constant for $M(r)$ such that it is the same as that for the dilaton field, i.e. $M$ as we have explained earlier in the previous section. Consequently, at "large-$r$" the exterior spacetime of the magnetic monopole configuration is represented by the metric
\begin{equation}
\begin{split}
ds^2=&-e^{(-\frac{1}{M^2r^2})}\left[1-\frac{2GM}{r}+\bigg(\frac{4\pi G}{e^2}+\frac{1}{M^2}\bigg)\frac{1}{r^2}\right]dt^2\\
&+\frac{1}{\left[1-\frac{2GM}{r}+(\frac{4\pi G}{e^2}+\frac{1}{M^2})\frac{1}{r^2}\right]}dr^2+r^2d\Omega^2_2
\end{split}
\end{equation}
where $M$ is an arbitrary mass parameter.

Before leaving this section we comment on the exact, analytic solution of the second combination of Einstein equations at large distance, Eg.(4.4).
First, notice that the Einstein equation in Eg.(4.4) can be recast into
\begin{equation}
\begin{split}
\frac{d}{dr}\left[M(r)e^{-\frac{1}{2M^2r^2}}\right]&=\left[4\pi\bigg(\frac{1}{8\pi GM^2}+\frac{1}{2e^2}\bigg)\frac{1}{r^2}\right]e^{-\frac{1}{2M^2r^2}}\\
&=\left[M'(r)+\frac{1}{M^2r^3}M(r)\right]e^{-\frac{1}{2M^2r^2}}
\end{split}
\end{equation}
which, upon integrating, becomes
\begin{equation}
M(r)e^{-\frac{1}{2M^2r^2}}-M(\infty)=4\pi\bigg(\frac{1}{8\pi GM^2}+\frac{1}{2e^2}\bigg)\int^r_\infty dr\frac{1}{r^2}e^{-\frac{1}{2M^2r^2}}.
\end{equation}
The integral on the right hand side, if we set $X\equiv\frac{1}{r}$, turns out to be the "error integral" which is related to the "incomplete gamma function". Namely the result is given by
\begin{equation}
\begin{split}
&M(r)=\left[M-2\pi\bigg(\frac{1}{e^2}+\frac{1}{4\pi GM^2}\bigg)\frac{M}{\sqrt{2}}\gamma\bigg(a=\frac{1}{2},\ \frac{1}{2M^2r^2}\bigg)\right]e^{\frac{1}{2M^2r^2}},\\
&A(r)=\left[1-\left\{\frac{2GM}{r}-\bigg(\frac{4\pi G}{e^2}+\frac{1}{M^2}\bigg)\frac{1}{r}\frac{M}{\sqrt{2}}\gamma\bigg(a=\frac{1}{2},\ \frac{1}{2M^2r^2}\bigg)\right\}e^{\frac{1}{2M^2r^2}}\right]^{-1}
\end{split}
\end{equation}
where we identified $M\equiv M(\infty)$ and $\gamma(a,z)$ is the incomplete gamma function defined by\\
\begin{equation*}
\begin{split}
\gamma(a,z)&=\int^z_0e^{-t}t^{a-1}dt\\
&=z^a\sum^\infty_{n=0}(-1)^{n}\frac{z^n}{n!(a+n)}
\end{split}
\end{equation*}
and related to the error integral by
\begin{equation*}
\begin{split}
erf(z)&=\frac{2}{\sqrt{\pi}}\int^z_0e^{-t^2}dt\\
&=\frac{1}{\sqrt{\pi}}\gamma\left(a=\frac{1}{2},\ z^2\right).
\end{split}
\end{equation*}
Finally, note that at large distance the infinite series expansion form of $\gamma(a,z)$ yields $\gamma(a=\frac{1}{2},\ \frac{1}{2M^2r^2})\simeq\frac{\sqrt{2}}{M}\frac{1}{r}$. Therefore, keeping the terms of order $\mathcal{O}(\frac{1}{r}),\ M(r)$ in Eq.(4.9) coincides with that in Eq.(4.5) as it should.\\\\
We have used
\begin{align*}
\int^r_\infty dr\frac{1}{r^2}e^{-\frac{1}{2M^2r^2}}&=-\int^X_0dXe^{-\frac{X^2}{2M^2}} \hspace{2cm} \bigg(X\equiv\frac{1}{r}\bigg)\\
&=-\sqrt{\frac{\pi}{2}}Merf\bigg(\frac{X}{\sqrt{2}M}\bigg)\\
&=-\frac{M}{\sqrt{2}}\gamma\bigg(a=\frac{1}{2},\ \frac{X^2}{2M^2}\bigg)
\end{align*}
"error integrals"
\[
  \begin{cases} 
   erf(z)=\frac{2}{\sqrt{\pi}}\int^z_0e^{-t^2}dt\\
   erf_c(z)=1-erf(z)=\frac{2}{\sqrt{\pi}}\int^\infty_ze^{-t^2}dt
  \end{cases}
\]
and their relations to gamma function and incomplete gamma function,
\[
  \begin{cases} 
   erf(z)=\frac{1}{\sqrt{\pi}}\gamma(a=\frac{1}{2},\ z^2)\\
   erf_c(z)=\frac{1}{\sqrt{\pi}}\Gamma(a=\frac{1}{2},\ z^2)
  \end{cases}
\]
where $\Gamma(a, z)$ and $\gamma(a,z)$ are defined by
\[
  \begin{cases} 
   \Gamma(a, z)=\int^\infty_ze^{-t}t^{a-1}dt\\
   \hspace{1.35cm}=z^{a-1}e^{-z}\sum^\infty_{n=0}(-1)^n\frac{(n-a)!}{(-a)!}\frac{1}{z^n}\\
   \gamma(a, z)=\Gamma(a)-\Gamma(a, z)\\
   \hspace{1.3cm}=\int^z_0e^{-t}t^{a-1}dt\\
   \hspace{1.3cm}=z^a\sum^\infty_{n=0}(-1)^n\frac{z^n}{n!(a+n)}
     \end{cases}
\]

\section{Positive definite monopole energy ("Bogonol'nyi bound")}
First, notice that the second combination of Einstein equations above, when integrated, leads to the total mass (energy) of the curved spacetime magnetic monopole, i.e.,
\begin{equation}
\begin{split}
&M(r)=\int^r_0dr'4\pi r'^2(-T^t_t)=4\pi\int^r_0dr'r'^2\rho_m(r')\\
&\text{thus}\\
&M\equiv M(\infty)=\int^\infty_0dr\ 4\pi r^2(-T^t_t)
\end{split}
\end{equation}
where $\rho_m(r)=(-T^t_t)=\left[\frac{1}{A}K+V\right]$ denotes the mass (energy) density of the system (i.e. Yang-Mill-Higgs system with dilaton field in curved spacetime).
This expression also shows that the "arbitrary mass parameter" M appearing in the exterior metric solution in Eq.(4.6) and Eq.(4.9) is clearly the total mass of our magnetic monopole in curved spacetime as it should be.

Now in this section, mainly following P. van Nieuwenhuizen et al. \cite{2} and K.Lee et al. \cite{3}, we shall show that our curved spacetime magnetic monopole mass also turns out to be "positive-definite" even in the presence of the dilaton field.
As was pointed out by P. van Nieuwenhuizen et al., unlike the case of YMH system in flat spacetime admitting the 't Hooft-Polyakov monopole solution (where the Lagrangian of the system is negative-definite), for the curved spacetime case like ours the Lagrangian of the non-metric sector $S_{non-metric}$ in Eq.(3.8) is not negative definite (of course the total Lagrangian $S_G+S_M$ is not negative-definite either). Therefore, although the Lagrangian is the negative of the energy for "static" systems like ours, we will not directly work with the Lagrangian $S_{non-metric}$ to show that our curved spacetime magnetic monopole has a positive-definite minimum energy. Instead, we will work with the expression for the energy (mass) of our curved spacetime monopole given by the second combination of Einstein equations in Eq.(4.2),
\begin{equation}
M'(r)=4\pi r^2[K(r)+V(r)]-8\pi GrM(r)K(r).
\end{equation}
This equation can be cast into the form,
\begin{equation}
\frac{d}{dr}\left[M(r)e^{-I(r)}\right]=\left\{4\pi r^2[K(r)+V(r)]\right\}e^{-I(r)}
\end{equation}
which, upon integration, yields
\begin{equation}
M(r)=\left\{4\pi\int^r_0dr'r'^2[K(r')+V(r')]e^{-I(r')}+M(0)e^{-I(0)}\right\}e^{I(r)}
\end{equation}
where $I(r)\equiv\int^\infty_rdr'8\pi Gr'K(r')$.\\ (The detailed proof of Eq.(5.4) is given in the \ref{inequ})\\\\
Thus the total energy (mass) of our curved spacetime monopole is given by
\begin{equation}
M=M(\infty)=4\pi\int^\infty_0dr\ r^2[K(r)+V(r)]e^{-I(r)}+M(0)e^{-I(0)}.
\end{equation}
It is already clear that if $M(0)\ge0$, then the total monopole energy $M$ is positive-definite since $K$ and $V$ are positive-definite. Further $M$ above satisfies the inequalities
\begin{equation}
M\ge\left\{4\pi\int^\infty_0dr\ r^2[K(r)+V(r)]+M(0)\right\}e^{-I(0)}\ge\frac{4\pi v}{e}e^{-I(0)} 
\end{equation}
where the first inequality is due to the positive-definiteness of $K$ whereas the second one follows from the Bogomol'nyi bound \cite{10}. And for a non-singular monopole solution, $M(0)=0$. The Eq.(5.6) above shows that the curved spacetime monopole mass $M$ has a positive-definite lower bound. Note also that for "static" systems like ours, a solution of classical field equations which maximizes the Lagrangian would minimizes the energy of the system since the energy is the negative of the Lagrangian. Therefore, if exists (actually we assume that it exists), a curved spacetime monopole solution (other than $h(r)=1,\ u(r)$=0 which is singular at $r=0$) is a "regular" localized soliton solution because its energy is finite and positive-definite.

Finally, since a curved spacetime solution $(w, u, h)$ of the classical field equations realizes the positive-definite minimum of $M$ (which is the functional of $w, u\ and\ h$), we have
\begin{equation}
M(w, u, h)\le M(w_0, u_0, h_0)\le4\pi\int^\infty_0dr\ r^2[K+V]\bigg|_{(w_0, u_0, h_0)}\le M_{\text{flat}}
\end{equation}
where $(w_0, u_0, h_0)$ is the flat spacetime classical solution which or course is different from its curved spacetime counterpart $(w, u, h)$ and $M_{\text{flat}}$ is the flat spacetime monopole mass. Clearly, this inequality reflects our general expectation that gravity tends to reduce the mass of a system because its overall effect is to bind the system.

\section{Transition to the monopole Black Hole}
In order to see if our curved spacetime monopole solution can actually make a transition to the "monopole black hole" solution, we investigate under what circumstances the monopole configuration collapse and eventually event horizons form. Thus we begin by considering the radial null geodesic. From the null condition $ds^2=-d\tau^2=0$, one gets the null geodesic equation
\begin{equation}
\left(\frac{dt}{dr}\right)^2=-\frac{g_{rr}}{g_{tt}}=e^{\frac{1}{M^2r^2}}\left[\frac{r^2}{r^2-2GMr+(\frac{4\pi G}{e^2}+\frac{1}{M^2})}\right].
\end{equation}
Thus future event horizons would form if $\left(\frac{dt}{dr}\right)\rightarrow\infty$ occurs or equivalently if $g_{rr}$ has poles.

Readily one can realize that if our curved spacetime monopole is a "large mass" monopole solution namely, if the total mass of our monopole $M$ is greater than the critical mass, viz.,
\begin{equation}
M\ge M_{\text{cr}}=\left[\left(\frac{2\pi}{Ge^2}\right)+\sqrt{\left(\frac{2\pi}{Ge^2}\right)^2+\frac{1}{G^2}}\right]^{\frac{1}{2}}
\end{equation}
then two event horizons form at
\begin{equation}
r_\pm=GM\pm\sqrt{G^2M^2-\left(\frac{4\pi G}{e^2}+\frac{1}{M^2}\right)}
\end{equation}
where $r_\pm$ denotes outer(+) and inner(-) event horizon respectively.\\
Note that since the mass of the monopole solution is $M\sim v/e$, comparison with the critical mass above shows that if the Higgs field vacuum expectation value $v$ (which is a free parameter of the theory) is comparable to or greater than the Plank mass, i.e. $v\ge M_{\text{Pl}}$ then the "large mass" monopole solution actually becomes a monopole black hole.

That $g_{rr}=A(r)$ indeed develops poles and thus event horizons form for large values of the dimensionless parameter $8\pi Gv^2$ has been illustrated numerically in the literature \cite{3} recently for the case of Einstein gravity. And in our case when the dilaton field is present, the metric has exactly the same behavior (of course to leading order) except that in the presence of the dilaton field, the critical mass for the transition to the monopole black hole to occur turns out to be greater than that in the case of Einstein gravity where $M_{\text{cr}}=\sqrt{\frac{4\pi}{Ge^2}}$ \cite{3}.

Now, since the large mass monopole solution does make a transition to the monopole black hole, it would be worth writing the exterior metric in terms of Kruskal coordinates. To do so, we first rewrite the exterior metric in terms of "null coordinate" as an intermediate step,
\begin{equation}
\begin{split}
ds^2&=-e^{-\frac{1}{M^2r^2}}\left[1-\frac{2GM}{r}+\left(\frac{4\pi G}{e^2}+\frac{1}{M^2}\right)\frac{1}{r^2}\right]dudv\\
&=-C(r)dudv
\end{split}
\end{equation}
where\quad  $u\equiv t-r_\ast$=const, $v\equiv t+r_\ast=$const denote outgoing and incoming null coordinates respectively and
\[
r_\ast\equiv e^{\frac{1}{2M^2r^2}}\left\{r+\left(\frac{r^2_+}{r_+-r_-}\right)\ln(r-r_+)(r-r_-)-\left(\frac{r^2_+-r^2_-}{r_+-r_-}\right)\ln(r-r_-)\right\}
\] 
is the generalized Regge-Wheeler tortoise coordinate written in terms of the outer$(r_+)$ and the inner$(r_-)$ event horizons defined earlier. (Here note that we keep only the leading terms in $r$ in the expression for $r_\ast$ since we are dealing with the exterior metric at large $r$.)

After following the usual procedure we arrive at the expression for the monopole black hole metric in terms of the Kruskal coordinates $(T, X, \theta, \phi)$
\begin{equation}
\begin{split}
ds^2=&\left(\frac{2r^2_+}{r_+-r_-}\right)^2\frac{1}{r^2}(r-r_-)^{\left(\frac{r^2_+-r^2_-}{r^2_+}\right)}\exp\left[-r\bigg/\left(\frac{r^2_+}{r_+-r_-}\right)\right](-dT^2+dX^2)\\
&+r^2d\Omega^2_2
\end{split}
\end{equation}
where again, we keep only the leading terms in $r$. And the relations between the old coordinates $(t, r)$ and the Kruskal coordinates $(T, X)$ are given by
\begin{equation}
\begin{split}
X^2-T^2&=\exp\left[r\bigg/\left(\frac{r^2_+}{r_+-r_-}\right)\right](r-r_-)^{-\left(\frac{r^2_+-r^2_-}{r^2_+}\right)}(r-r_+)(r-r_-),\\
\frac{T}{X}&=\tanh\left[e^{-\frac{1}{2M^2r^2}}\left(\frac{r_+-r_-}{2r^2_+}\right)t\right]
\end{split}
\end{equation}
since the Kruskal coordinates are defined by
\begin{equation}
\begin{cases}
T=\exp\left[e^{-\frac{1}{2M^2r^2}}\left(\frac{r_+-r_-}{2r^2_+}\right)r_\ast\right]\sinh\left[e^{-\frac{1}{2M^2r^2}}\left(\frac{r_+-r_-}{2r^2_+}\right)t\right],\\
X=\exp\left[e^{-\frac{1}{2M^2r^2}}\left(\frac{r_+-r_-}{2r^2_+}\right)r_\ast\right]\cosh\left[e^{-\frac{1}{2M^2r^2}}\left(\frac{r_+-r_-}{2r^2_+}\right)t\right].
\end{cases}
\end{equation}
Now several comments are in order concerning the properties of our monopole black hole and the structure of tis event horizons.

Above all, it turned out that in our theory where the dilaton field is present, the critical mass for the transition to the monopole black hole is somewhat greater than that in the case of Einstein gravity.

Now, just as the usual Reissner-Nordstr{\"o}m metric in general relativity, monopole black hole in stringy gravity metric has both inner and outer horizons (to leading order at large $r$). However, due to the presence of the dilaton field that leads to the larger critical mass, these two horizons of our monopole black hole got closer to each other than the usual Reissner-Nordstr{\"o}m black hole horizons(namely, $r^{\text{ours}}_-> r^{\text{RN}}_-,\  r^{\text{ours}}_+<r^{\text{RN}}_+$). And the spacetime geometry is not singular at either of these two event horizons as is obvious in the form of our monopole black hole metric expressed in terms of Kruskal coordinates in Eq.(6.5).

Next, it may be with noting that the inner horizon is unstable whereas the other horizon is stable in that non-spherically symmetric (i.e. anisotropic) perturbations tend to blow up on the inner event horizon as is well-known in the case of usual Reissner-Nordstr{\"o}m metric \cite{11}.

Finally, notice that the physical distance to the event horizon is infinite although it can be traversed in finite proper time for our monopole black hole, namely
\begin{equation}
L(r_H)=\int^{r_H}_0\sqrt{g_{rr}}dr
\end{equation}
diverges since $\sqrt{g_{rr}}$ behaves as $r/(r-r_H)$ near the horizon. This fact, then, implies that all the non-trivial field configurations representing the 'entire' structure of our non-Abelian monopole take place only within the event horizon in accordance with the "no-hair" theorems of black holes \cite{12}.\\
Now we turn to the evolution of our monopole black hole in stringy gravity.
The relationship between the old coordinates $(t, r)$ and the Kruskal coordinates $(T, X)$ in Eq.(6.6) reveals that the spatial coordinate $r$ actually is an implicit function of $T$ which is a "global time" coordinate. Therefore just like the usual Reissner-Nordstr{\"o}m metric, monopole black hole in stringy gravity metric does have time dependence and hence the spacetime is not static globally. This characteristic of the metric, then, leads to the black hole evaporation via the emission of particles, namely the "Hawking radiation" \cite{13}. And probably the quickest way to find out the Hawking temperature $T_H$ would be to read it off from the periodicity in time coordinate of the Euclidean section.

Here, however, we shall take the usual method to compute the Hawking temperature \cite{14}. Namely, in terms of the "surface gravity" $\kappa$, the Hawking temperature is written as
\begin{equation}
T_H=\frac{\kappa}{2\pi}.
\end{equation}
Thus the track of finding out the Hawking temperature reduces to the calculation of the surface gravity.
And it is known \cite{14} that the surface gravity $\kappa$ is related to the metric written in terms of the null coordinates as in Eq.(6.4) by
\begin{equation}
\kappa=\frac{1}{2}\frac{\partial C}{\partial r}\bigg|_{r=r_+}. 
\end{equation}
For monopole black hole in stringy gravity with the exterior metric given by Eq.(6.4), the surface gravity is found to be
\begin{equation}
\kappa=e^{-\frac{1}{M^2r^2_+}}\left(\frac{r_+-r_-}{2r^2_+}\right)
\end{equation}
which then leads to the Hawking temperature
\begin{equation}
\begin{split}
T_H=&\frac{1}{2\pi}e^{-\frac{1}{M^2r^2_+}}\left(\frac{r_+-r_-}{2r^2_+}\right)\\
=&\frac{1}{2\pi G}\exp\left[-\frac{1}{G^2M^2\left[M+\sqrt{M^2-\frac{1}{G^2M^2}-\left(M^2_{cr}-\frac{1}{G^2M^2_{cr}}\right)}\right]^2}\right]\\
&\times\frac{\sqrt{M^2-\frac{1}{G^2M^2}-\left(M^2_{cr}-\frac{1}{G^2M^2_{cr}}\right)}}{\left[M+\sqrt{M^2-\frac{1}{G^2M^2}-\left(M^2_{cr}-\frac{1}{G^2M^2_{cr}}\right)}\right]^2}
\end{split}
\end{equation}
where $M_{cr}$ is as defined in Eq.(6.2).
Now, as our monopole black hole loses its mass by emitting particles, its Hawking temperature given above slowly increases, reaching a maximum temperature and then falls rapidly to zero as it approaches the extremal black hole solution (i.e. maximally charged hole) $M\rightarrow M_{cr}$. In other words, our magnetically charged monopole black hole in stringy gravity does not evaporate completely. This is, in the sense of "cosmic censorship hypothesis", fortunate because further evaporation after reaching $M=M_{cr}$ would lead to a naked singularity. Thus the extreme dilaton black hole solutions are stable end points of the Hawking radiation.

\begin{figure}[h!]
\includegraphics[width=10cm, angle=0]{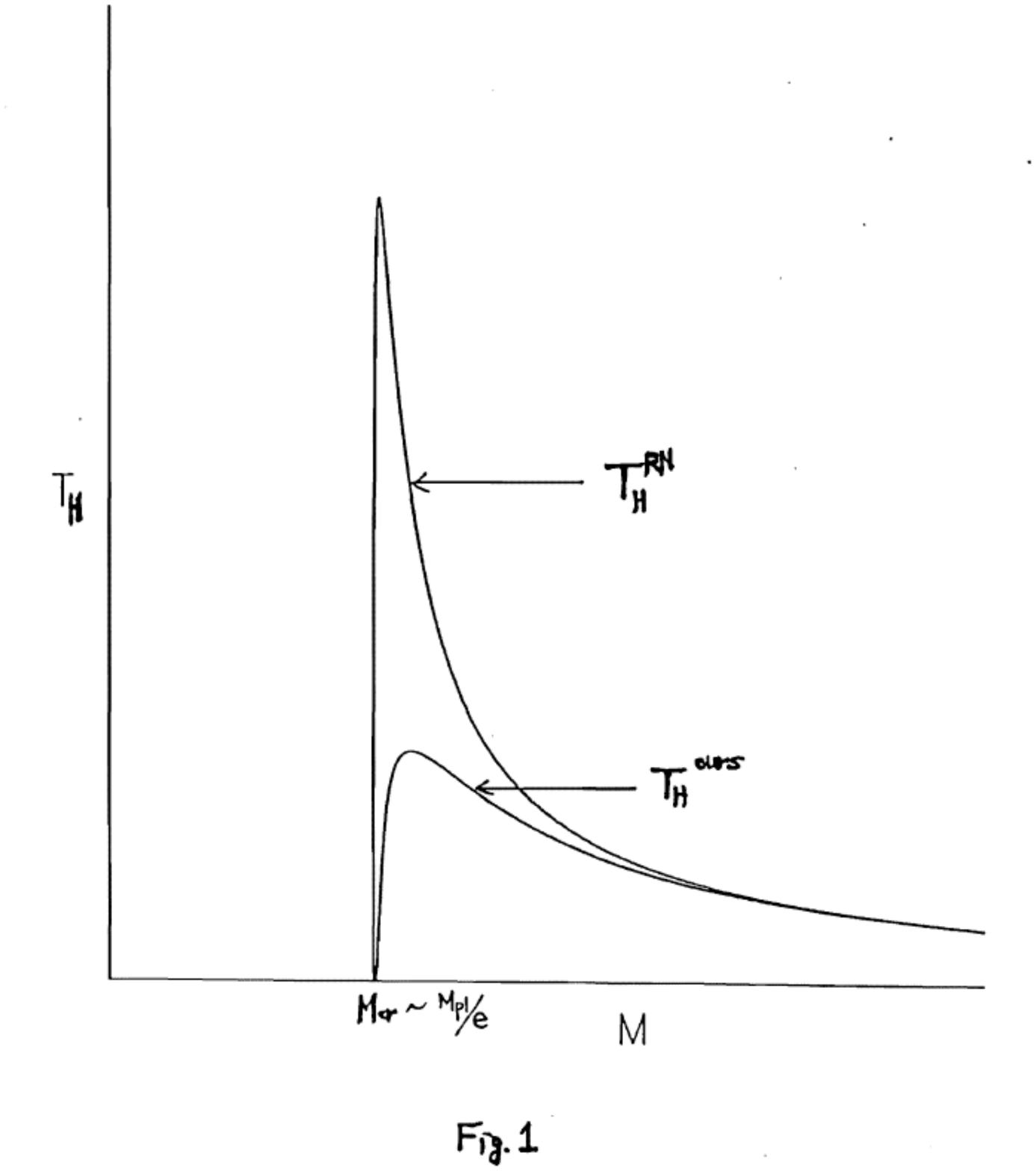}
\caption{The profile for the Hawking temperature of the monopole black hole. A "mass dependent extra attractive interaction" in addition to that of the standard Einstein gravity generated by the dilaton field exhibits the regularization of the singular behavior of the Hawking temperature that blows up at a critical point which indeed happens for the case of the self-gravitating RN black hole in the absence of the dilaton field.}
\label{fig:1}
\end{figure}

These general features of the black hole evolution stated above are essentially the same as those in the case of monopole black holes in Einstein gravity. There is, however, an important difference for monopole black hole in stringy gravity, the Hawking temperature is generally lower (namely for a given $M,\  T^{ours}_H<T^{RN}_H)$ and goes more rapidly to zero than their Einstein gravity's counterpart as they lose mass. The plot of Hawking temperature versus black hole mass for monopole black hole in stringy gravity and for Einstein gravity monopole black holes is given in Fig.\ref{fig:1}. This behavior of monopole black hole in stringy gravity seems to imply that the presence of the dilaton field introduces an extra attraction interaction.
Earlier in this section, it is found that the critical mass for the transition to monopole black holes is somewhat greater than that in Einstein gravity case. And it can be regarded as another implication that the nature of new interaction that the dilaton introduces is an extra attractive force.

Here, we need careful analysis regarding the nature of new interaction the dilaton seems to introduce.
Namely, it is important to recognize that the two observed properties, i.e., the grater critical mass for the transition to monopole black holes and the generally lower Hawking temperature which also falls more rapidly to zero, should not be regarded as being inconsistent. Rather, these two features imply that the nature of new interaction is an extra attractive force. 
In fact, whenever an additional field with canonical kinetic term and non-trivial asymptotic behavior (such as the Yang-Mills field or the dilaton field in our system) is introduced into the theory, its kinetic energy always makes an extra positive-definite contribution to the energy density of the system $\rho_m=-T^t_t$ in Eq.(4.1) and (4.2) regardless of the specifics of the field added (such as the dilaton charge). Now this increase in the energy density of the system, when translated into the solution of Einstein equation, leads to the greater critical mass for the transition to the monopole black holes than for the case without the additional field as is manifest from Eq.(3.8) and (4.2). Thus the greater critical mass in the presence of the dilaton field is a trivial, expected consequence which has little to do with the details of the nature of some new interaction the dilaton might introduce into the system. Rather, it is the features of the dilaton field such as its non-zero dilaton charge that would actually exhibit the nature of the new interaction introduced by the dilaton.

And of course the subsequent evolution such as the Hawking evaporation after the monopole became a black hole would reveal the effects of the new interaction.
Therefore, in terms of the characteristics of the dilaton charge found in the previous section, we shall explain the behavior of the Hawking temperature observed above. Then this analysis will allow us to identify the nature of the new interaction due to the dilaton. First, the fact that the dilaton equation of motion in Eq.(3.5b) is a source-free Poisson equation tells us that the dilaton field is essentially massless classically and hence introduces a long rage forces. Next, recall from Eq.(3.7) that the dilaton charge is 'negative" and inversely proportional to the mass of the monopole, $D=-1/M$. Obviously, the new interaction introduced by the dilaton field would be directly proportional to this "dilaton charge" $D$. Therefore interpreted in terms of the character of this dilaton charge, $D$, the correct identification of the new interaction would be the "mass-dependent extra attractive interaction".

That is, as the mass of the monopole black hole increase the extra attractive force introduced by the dilaton field decreases whereas as the mass decreases the extra attraction increases in magnitude. And indeed, the effect of this extra attractive force introduced by the dilaton field is exactly realized in the Hawking evaporation of monopole black hole in stringy gravity as is manifest in the behavior of the Hawking temperature plotted in Fig.\ref{fig:1}. in other words, as the monopole black hole loses its mass via the particle emission the extra attractive force introduced by the dilaton field grows and as a result its Hawking temperature falls and hence the Hawking evaporation terminates faster than in the case of Einstein gravity.

To conclude, the presence itself of the dilaton field automatically leads to the greater critical mass for the transition to monopole black holes, but the real nature of the new interaction the dilaton introduces is essentially determined by the dilaton charge and it turns out to be the "mass-dependent extra attractive interaction".

\section{Conclusions}
Now we summarize our results. We have studied the possible transition of "large mass" magnetic monopole solution to the monopole black hole solution in the Yang-Mills-Higgs theory with SSB coupled to the 4-dimensional low energy effective theory of superstring with only dilatonic sector. Our main interest was in the effects of the dilaton field on the nature of classical black hole solutions of our theory. For the metric solution of Einstein equations describing the exterior spacetime of the monopole black hole, first the critical mass for the dilaton monopole black hole is found to be greater than that for the monopole black hole in Einstein gravity. Second, it has been found that as the monopole black hole loses its mass via Hawking radiation and thus moves toward the "extremal" black hole (i.e. $M\rightarrow M_{cr}$) the Hawking temperature of monopole black hole in stringy gravity is generally lower and falls more rapidly to zero than in the case of Einstein gravity. As analyzed in detail earlier, these properties of the monopole black hole in stringy gravity solutions lead us to conclude that the presence of the dilaton field appears to introduce an additional attraction force which is inversely proportional to the mass of the monopole black hole. This observation concerning the effects of the presence of the dilaton field on the monopole black hole solution is in accordance with the known properties of the charged black hole solutions in low energy string theory that in string theory the dilaton contributes an extra attractive force to the magnetically charged black hole solution \cite{5}.

Now we comment on one more aspect of the system we considered. It is straightforward to see that an exact, albeit singular, monopole solution of Wu-Yang type exists again independently of the gravitational fields (metric field and dilaton field) in the absence of the Higgs scalar field in our theory. And then it follows that for this Wu-Yang monopole solution in curved spacetime, the dilaton solution and the metric solution (at large distances) remain the same as in the case of 't Hooft-Polyakov monopole solution in curved spacetime. Thus the exterior spacetime of the Wu-Yang monopole configuration is described by the metric of almost Reissner-Nordstr{\"o}m type given in Eq.(4.6). However, the transition of this Wu-Yang monopole solution to a black hole solution is unclear (in fact, unlikely) because there is no such free parameter in the theory as the Higgs field vacuum expectation valve $v$ by which the transition to black holes can be probed.

Next, we stress the possible cosmological implications of the non-Abelian monopole black holes in general. As mentioned earlier in the introduction, non-Abelian monopole black holes like the one explored in the present work may help explain away two cosmological puzzles at a single stroke.
Namely, if "large mass" magnetic monopoles do make transitions to monopole black holes as they form, it is indeed no surprise that non-Abelian monopoles have never been seen (although the inflationary universe scenario provides yet another explanation for this puzzle).
This is because, as we have observed earlier in this work, all the non-trivial field configurations such as the non-vanishing $SU(2)$-magnetic field strength by which an asymptotic observer can identify them with non-Abelian monopoles are completely trapped within the event horizon. Thus there would be simply no way for an asymptotic observer to identify these black holes with non-Abelian monopoles he might have been searching for (Of course this behavior of monopole black hole is connected to the "no-hair" theorems according to which a static black hole does not reveal any non-trivial field configurations outside the event horizon.)

Secondly since in principle arbitrary number of non-Abelian magnetic monopoles can be produced in those unified particle theories \cite{9} and since they are quite 'heavy' $(M_{mon} \sim v/e \sim (\frac{1}{e})M_{pl} \sim (\frac{1}{e})10^{-5}g$ where $e \ll 1$ in the weak gauge coupling limit as is usually the case) and of course completely 'dark' after they become monopole black holes, non-Abelian monopole black holes may be a dominating component of "cold dark matter" which is believed to exist to reconcile with the missing mass problem of the universe.

Aside from these significant roles played by the non-Abelian monopole black holes in resolving two major cosmological puzzles, they also may exhibit some interesting features. For instance, these non-Abelian monopole black holes are "mini black holes" and essentially would look like elementary particles since their size is as small as $R_{mon} \sim 1/ev \sim (\frac{1}{e})l_{pl} \sim (\frac{1}{e})10^{-33}$cm. (It may seem rather contradictory that their size is finite and this small while the physical horizon radius is infinite as pointed out before. This is because the coordinate distance is what we measure and the physical distance is what it feels like.) This size is, in the limit of weak gauge coupling, $e\ll 1$, small enough to view them as entities more like elementary particles than extended objects but still much larger than the Plank length $l_{pl}$ justifying the neglect of quantum gravity effects.

After all, non-Abelian monopole black holes seem to deserve further careful investigations not only for their interesting theoretical aspects but also for the significant cosmological implications they may have.

%\section*{References}

\appendix
\section{Low energy effective theory of the bosonic part of superstring compactified to D-dims.}
\[
\begin{split}
S=\int d^Dx\sqrt{g}e^{-2\Phi(x)}&\left\{\frac{1}{2k^2}\left[R+4(\nabla\Phi)^2-\frac{1}{4}F^2_{(2)}-\frac{1}{2(D-2)!}F^2_{(D-2)}\right.\right.\\
&\left.\left.-\frac{1}{12}H^2_{\mu\nu\lambda}-\Lambda\right]\right\}
\end{split}
\]
$\Downarrow [g_{\mu\nu}(x)=e^{(\frac{4}{D-2})\Phi(x)}\tilde{g}_{\mu\nu}(x)]$\\\\
\[
\begin{split}
S=\int d^Dx\sqrt{g}&\left\{\frac{1}{2k^2}\left[R-\frac{4}{(D-2)}(\nabla\Phi)^2-\frac{1}{4}e^{-(\frac{4}{D-2})\Phi}F^2_{(2)}\right.\right.\\
&\left.\left.-\frac{1}{2(D-2)!}e^{-4(\frac{D-3}{D-2})\Phi}F^2_{(D-2)}-\frac{1}{12}e^{-(\frac{8}{D-2})\Phi}H^2_{\mu\nu\lambda}-e^{(\frac{4}{D-2})\Phi}\Lambda\right]\right\}
\end{split}
\]
where
\[
\begin{cases}
k^2\equiv8\pi G\\
F_{(2)}: \text{gauge field strength tensor 2-form}\\
F_{(D-2)}: \text{gauge field strength tensor (D-2)-from}\\
H_{\mu\nu\lambda}: \text{antisymmetric field strength tensor 3-form}
\end{cases}
\]
$H=dB+\omega^0_{3L}-\omega^0_{3Y}$
\[
\begin{cases}
B: \text{potential 2-form}\\
\omega^0_{3L} = \text{Tr}[\omega\wedge d\omega+\frac{2}{3}\omega\wedge\omega\wedge\omega]\\
\cdots\cdots\text{Chern-Simons form for the Lorentz spin connection}\\
\omega^0_{3Y}=\text{Tr}[A\wedge dA+\frac{2}{3}A\wedge A\wedge A]\\
\cdots\cdots\text{Chern-Simons form for the gauge connection.}
\end{cases}
\]
Now, consider the Weyl rescaling of the action that represents our theory,
\begin{align*}
S&=\int d^4x\sqrt{g}\left\{e^{-2\Phi(x)}\frac{1}{2k^2}\left[R+4(\nabla\Phi)^2\right]+\mathcal{L}_M\right\}\\
&=\int d^4x\sqrt{g}\left\{e^{-2\Phi(x)}\frac{1}{2k^2}\left[R+4(\nabla\Phi)^2\right]-\frac{1}{4}(F^a_{\mu\nu})^2-\frac{1}{2}(D_\mu\phi^a)^2-U(\phi^a\phi^a)\right\}.
\end{align*}
where   \quad$U(\phi^a\phi^a)=\frac{\lambda}{4}(\phi^a\phi^a-v^2)^2,\ k^2\equiv8\pi G$.\\
Under the Weyl rescaling of gravitational fields (i.e. the metric and the dilaton field),
\begin{align*}
&g_{\mu\nu}(x)=\Omega^2(x)\tilde{g}_{\mu\nu}(x)=e^{2\Phi(x)}\tilde{g}_{\mu\nu}(x),\\
&\Phi(x)=\Omega^{-d}(x)\tilde{\Phi}(x)=\tilde{\Phi}(x)
\end{align*}
(where the scale dimension of $\Phi(x)$ field is $d=0$ since $\Phi(x)$ is dimensionless in the gravity action written in "sigma model" form)\\
it follows that 
\begin{align*}
\sqrt{g}&=\sqrt{\text{det}g_{\mu\nu}}=\Omega^4(x)\sqrt{\tilde{g}}=e^{4\Phi(x)}\sqrt{\tilde{g}}\\
g^{\mu\nu}(x)&=\Omega^{-2}(x)\tilde{g}^{\mu\nu}(x)=e^{-2\Phi(x)}\tilde{g}^{\mu\nu}(x)\\
R(g)&=\Omega^{-2}(x)\left[\tilde{R}(\tilde{g})-6\Omega^{-1}(x)\square\Omega(x)\right]\\
&=\Omega^{-2}\tilde{R}(\tilde{g})-6\Omega^{-3}\square\Omega
\end{align*}
Then,
\begin{align*}
\begin{split}
S=&\int d^4x\sqrt{\tilde{g}}\Omega^4(x)\left\{\Omega^{-2}\frac{1}{2k^2}\left[\left\{\Omega^{-2}\tilde{R}-6\Omega^{-3}\frac{1}{\sqrt{\tilde{g}}}\partial_\mu(\sqrt{\tilde{g}}\tilde{g}^{\mu\nu}\partial_\nu\Omega)\right\}+4\Omega^{-2}\tilde{g}^{\mu\nu}\partial_\mu\tilde{\Phi}\partial_\nu\tilde{\Phi}\right]\right.\\
&\left.-\frac{1}{4}\Omega^{-4}\tilde{g}^{\mu\alpha}\tilde{g}^{\nu\beta}F^a_{\mu\nu}F^a_{\alpha\beta}-\frac{1}{2}\Omega^{-2}\tilde{g}^{\mu\nu}
(D_\mu\phi^a)(D_\nu\phi^a)-U(\phi^a\phi^a)\right\}\\
=&\int d^4x\sqrt{\tilde{g}}\left\{\frac{1}{2k^2}\left[\tilde{R}-6\Omega^{-1}\frac{1}{\sqrt{\tilde{g}}}\partial_\mu(\sqrt{\tilde{g}}\tilde{g}^{\mu\nu}\partial_\nu\Omega)+4\tilde{g}^{\mu\nu}\partial_\mu\tilde{\Phi}\partial_\nu\tilde{\Phi}\right]\right.\\
&\left.-\frac{1}{4}\tilde{g}^{\mu\alpha}\tilde{g}^{\nu\beta}F^a_{\mu\nu}F^a_{\alpha\beta}-\frac{1}{2}\Omega^{2}\tilde{g}^{\mu\nu}
(D_\mu\phi^a)(D_\nu\phi^a)-\Omega^4U(\phi^a\phi^a)\right\}\\
=&\int d^4x\sqrt{\tilde{g}}\left\{\frac{1}{2k^2}\left[\tilde{R}+4\tilde{g}^{\mu\nu}\partial_\mu\tilde{\Phi}\partial_\nu\tilde{\Phi}-6\Omega^{-2}\tilde{g}^{\mu\nu}(\partial_\mu\Omega)(\partial_\nu\Omega)\right]\right.\\
&\left.-\frac{1}{4}\tilde{g}^{\mu\alpha}\tilde{g}^{\nu\beta}F^a_{\mu\nu}F^a_{\alpha\beta}-\frac{1}{2}\Omega^{2}\tilde{g}^{\mu\nu}
(D_\mu\phi^a)(D_\nu\phi^a)-\Omega^4U(\phi^a\phi^a)\right\}\\
=&\int d^4x\sqrt{\tilde{g}}\left\{\frac{1}{2k^2}\left[\tilde{R}-2\tilde{g}^{\mu\nu}\partial_\mu\tilde{\Phi}\partial_\nu\tilde{\Phi}\right]\right.\\
&\left.-\frac{1}{4}\tilde{g}^{\mu\alpha}\tilde{g}^{\nu\beta}F^a_{\mu\nu}F^a_{\alpha\beta}-\frac{1}{2}e^{2\tilde{\Phi}(x)}\tilde{g}^{\mu\nu}
(D_\mu\phi^a)(D_\nu\phi^a)-e^{4\tilde{\Phi}(x)}U(\phi^a\phi^a)\right\}
\end{split}
\end{align*}
where we used

\begin{align*}
\int& d^4x\sqrt{\tilde{g}}\left\{\frac{1}{2k^2}\left[-6\Omega^{-1}\frac{1}{\sqrt{\tilde{g}}}\partial_\mu(\sqrt{\tilde{g}}\tilde{g}^{\mu\nu}\partial_\nu\Omega)\right] \right\}\\
=&\int d^4x\left\{\frac{1}{2k^2}(-6)\Omega^{-1}\partial_\mu(\sqrt{\tilde{g}}\tilde{g}^{\mu\nu}\partial_\nu\Omega)\right\}\\
=&\int d^4x\left\{\frac{1}{2k^2}(-6)\partial_\mu(\Omega^{-1}\sqrt{\tilde{g}}\tilde{g}^{\mu\nu}\partial_\nu\Omega)\right\}\\
&-\int d^4x\sqrt{\tilde{g}}\left\{\frac{1}{2k^2}(-6)\tilde{g}^{\mu\nu}(\partial_\mu\Omega^{-1}\partial_\nu\Omega)\right\}\\
=&-\int d^4x\sqrt{\tilde{g}}\left\{\frac{1}{2k^2}6\Omega^{-2}\tilde{g}^{\mu\nu}(\partial_\mu\Omega)(\partial_\nu\Omega)\right\}
\end{align*}
and $\Omega(x)=e^{\Phi(x)}$\ thus\  $(\partial_\mu\Omega)=e^\Phi(\partial_\mu\Phi)$.

Thus, after this Weyl rescaling, the actin takes the form in "Einstein conformal" gauge,
\[S=\int d^4x\sqrt{\tilde{g}}\left\{\frac{1}{2k^2}[\tilde{R}-2(\nabla\tilde{\Phi})^2]-\frac{1}{4}(F^a_{\mu\nu})^2-\frac{1}{2}e^{2\tilde{\Phi}}(D_\mu\phi^a)^2-e^{4\tilde{\Phi}}U(\phi^a\phi^a)\right\}.\]
Note, first, that the dilaton field $\Phi(x)$, which used to be a ghost field with negative definite norm, now takes the kinetic term of the canonical form and hence turns into a physical scalar field (with positive definite norm) in the Einstein conformal gauge after the Weyl rescaling of metric and dilaton field.

We will work in this "Einstein conformal" gauge in which the Einstein-Hilbert action rather than some multiple of it by a non-linear realization of the dilaton field appears. We choose this conformal gauge because the physical interpretation such a s the behavior of horizons and black hole thermodynamics are more conveniently discussed in this gauge.

Then our theory is represented by the action,
\[
S=\int d^4x\sqrt{g}\left\{\frac{1}{16\pi G}\left[R-2(\nabla\Phi)^2\right]-\frac{1}{4}(F^a_{\mu\nu})^2-\frac{1}{2}e^{2\Phi}(D_\mu\phi^a)^2-e^{4\Phi}U(\phi^a\phi^a)\right\}
\]
where we dropped tilde.

\section{Hawking Radiation of the monopole black hole}

In order to study the evaporation of our monopole black hole, we first infer its Hawking temperature $T_H$ from the periodicity of the Euclidean time coordinate by writing the monopole black hole metric in terms of (generalized) Kruskal coordinates as follows:\\
The metric describing the exterior spacetime to the monopole configuration is,
\[
\begin{split}
ds^2=-e^{\left(-\frac{c^2}{r^2}\right)}\left[1-\frac{2GM}{r}+(4\pi GQ^2+c^2)\frac{1}{r^2}\right]dt^2\\+\frac{1}{\left[1-\frac{2GM}{r}+(4\pi GQ^2+c^2)\frac{1}{r^2}\right]}dr^2+r^2d\Omega^2_2.
\end{split}
\]
Now, in order to transform eventually to Kruskal coords., we look for the "radial null-geodesic" using the (radial) null condition, $ds^2=-d\tau^2=0$
\begin{align*}
&\left(\frac{dt}{dr}\right)^2=\frac{1}{e^{\left(-\frac{c^2}{r^2}\right)}\left[1-\frac{2GM}{r}+(4\pi GQ^2+c^2)\frac{1}{r^2}\right]^2}\\
&\left(\frac{dt}{dr}\right)=\pm e^{\left(\frac{c^2}{2r^2}\right)}\left[\frac{r^2}{r^2-2GMr+(4\pi GQ^2+c^2)}\right]
\end{align*}
Here, note that when event horizons occur and the black hole forms, $\left(\frac{dt}{dr}\right)\rightarrow\infty$ as $t\rightarrow+\infty.$
Thus the future event horizons can be found by setting\\
 $\left(\frac{dt}{dr}\right)\rightarrow\infty \Rightarrow Cr^2+Br+A=0.$\\\\
where\\\\  
$\begin{cases}
A\equiv(4\pi GQ^2+c^2)\\
B\equiv-2GM\\
C\equiv1
\end{cases}$\\\\
Now, for two roots to exist, $B^2-4AC > 0\ (\text{i.e.}\ G^2M^2-(4\pi GQ^2+c^2) > 0 )$\\
and $r=\frac{1}{2c}\left[-B\pm\sqrt{B^2-4AC}\right]\equiv r_{\pm}$\\
$\left(\text{i.e.}\ r_{\pm}=\left[GM\pm\sqrt{G^2M^2-(4\pi GQ^2+c^2)}\right]\right)$
\begin{align*}
t&=\pm\int dr\ \frac{1}{e^{\left(-\frac{c^2}{2r^2}\right)}\left[1-\frac{2GM}{r}+(4\pi GQ^2+c^2)\frac{1}{r^2}\right]}\\
&=\pm\int dr\ e^{\left(\frac{c^2}{2r^2}\right)}\frac{r^2}{\left[r^2-2GMr+(4\pi GQ^2+c^2)\right]}\cdots\cdots\star\\
&\equiv\pm\int dr\ f'(r)g(r)
\end{align*}
Now, we use the 'integration by part', 
\begin{align*}
f'(r)&\equiv\frac{r^2}{Cr^2+Br+A}\\
g(r)&\equiv e^{\left(\frac{c^2}{2r^2}\right)}\\
f(r)&=\int dr\frac{r^2}{(Cr^2+Br+A)}\\
&=\frac{r}{C}-\frac{B}{2C^2}\ln(Cr^2+Br+A)+\frac{B^2-2AC}{2C^2}\underbrace{\int dr\frac{1}{(Cr^2+Br+A)}}_{=\frac{1}{\sqrt{\triangle}}\ln\left( \frac{B+2Cr-\sqrt{\triangle}}{B+2Cr+\sqrt{\triangle}} \right)}\\
&=\frac{r}{C}-\frac{B}{2C^2}\ln(Cr^2+Br+A)\\
&\hspace{3cm}+\frac{1}{2C^2}\frac{B^2-2AC}{\sqrt{B^2-4AC}}\ln\left( \frac{B+2Cr-\sqrt{B^2-4AC}}{B+2Cr+\sqrt{B^2-4AC}} \right)\\
g'(r)&=\left(-\frac{c^2}{r^3}\right)e^{\left(\frac{c^2}{2r^2}\right)}
\end{align*}
where\\\\
$\begin{cases}
A\equiv(4\pi GQ^2+c^2)\\
B\equiv-2GM\\
C\equiv1
\end{cases}$\\\\
$\triangle\equiv B^2-4AC>0\ (\text{for two event horizons to exist})$\\\\
$\Rightarrow$ Thus,
\begin{align*}
t=&\pm\left[f(r)g(r)-\int dr f(r)g'(r)\right]\\
=&\pm\left[e^{\left(\frac{c^2}{2r^2}\right)}\left\{\frac{r}{C}-\frac{B}{2C^2}\ln(Cr^2+Br+A)\right.\right.\\
&\hspace{3cm}\left.\left.+\frac{1}{2C^2}\frac{B^2-2AC}{\sqrt{B^2-4AC}}\ln\left(\frac{B+2Cr-\sqrt{B^2-4AC}}{B+2Cr+\sqrt{B^2-4AC}}\right)\right\}\right.\\
&\left.+\int dr\ e^{\left(\frac{c^2}{2r^2}\right)}\frac{c^2}{r^3}\left\{\frac{r}{C}-\frac{B}{2C^2}\ln(Cr^2+Br+A)\right.\right.\\
&\hspace{3cm}\left.\left.+\frac{1}{2C^2}\frac{B^2-2AC}{\sqrt{B^2-4AC}}\ln\left(\frac{B+2Cr-\sqrt{B^2-4AC}}{B+2Cr+\sqrt{B^2-4AC}}\right)\right\}\right]\\
&+\text{const.}
\end{align*}
Note, however, that our metric solution describes the "exterior spacetime" to the monopole configuration and hence essentially valid for large $r$.
Therefore in the above equation for the radial null geodesic, the second term, which is of order $\mathcal{O}\left(e^{\left(\frac{1}{r^2}\right)}\frac{1}{r}\right)$, is certainly negligible compared to the first term. (In fact, this can be thought of as an approximation in which the term $e^{\left(\frac{c^2}{2r^2}\right)}$ in the integrand of integral $\star$ is regarded as an almost constant in the large - $r$ region and thus gets out of the integral sign.)

Thus we have the equation for the radial null geodesic,

\begin{align*}
t=&\pm\left[e^{\left(\frac{c^2}{2r^2}\right)}\left\{\frac{r}{C}-\frac{B}{2C^2}\ln(Cr^2+Br+A)\right.\right.\\
&\hspace{2cm}\left.\left.+\frac{1}{2C^2}\frac{B^2-2AC}{\sqrt{B^2-4AC}}\ln\left(\frac{B+2Cr-\sqrt{B^2-4AC}}{B+2Cr+\sqrt{B^2-4AC}}\right)+\mathcal{O}\left(\frac{1}{r}\right)\right\}\right]\\
&+\text{const.}\\
\simeq&\pm\left[e^{\left(\frac{c^2}{2r^2}\right)}\left\{r+GM\ln(r^2-2GMr+[4\pi GQ^2+c^2])\right.\right.\\
&\left.\left.+\frac{2G^2M^2-(4\pi GQ^2+c^2)}{2\sqrt{G^2M^2-(4\pi GQ^2+c^2)}}\ln\left[\frac{r-GM-\sqrt{G^2M^2-(4\pi GQ^2+c^2)}}{r-GM+\sqrt{G^2M^2-(4\pi GQ^2+c^2)}}\right]\right\}\right]\\
&+\text{const.}\\
\equiv&\pm r_\ast+\text{const.} \hspace{2cm}\therefore t\mp r_\ast=\text{const.}
\end{align*}
where we introduced, "(generalized) Regge-Wheeler tortoise" coordinate which can be rewritten in terms of outer and inner event horizons
\[r_{\pm}\equiv GM\pm\sqrt{G^2M^2-(4\pi GQ^2+c^2)}\]
as
\begin{align*}
r_\ast\equiv& e^{\left(\frac{c^2}{2r^2}\right)}\left\{r+\left(\frac{r_++r_-}{2}\right)\ln(r-r_+)(r-r_-)+\frac{1}{2}\left(\frac{r^2_++r^2_-}{r_+-r_-}\right)\ln\left(\frac{r-r_+}{r-r_-}\right)\right\}\\
=&e^{\left(\frac{c^2}{2r^2}\right)}\left\{r+\left(\frac{r^2_+}{r_+-r_-}\right)\ln(r-r_+)(r-r_-)-\left(\frac{r^2_+-r^2_-}{r_+-r_-}\right)\ln(r-r_-)\right\}
\end{align*}
then
\begin{align*}
r_+\equiv GM+\sqrt{G^2M^2-(4\pi GQ^2+c^2)}\cdots \text{"outer" event horizon},\\
r_-\equiv GM-\sqrt{G^2M^2-(4\pi GQ^2+c^2)}\cdots \text{"inner" event horizon}.
\end{align*}
\begin{table}[h!]
    \begin{tabular}{l|c}  $r$ & $r_\ast$\\ 
      \hline
      $\infty$ & $\infty$\\
      $r_{\pm}$ & -$\infty$\\
  \end{tabular}
\end{table}\\
(Note that $r_\ast=-\infty$ is not necessarily true for the "inner" horizon $r=r_-.$)\\\\
Now, define "null coordinates" $(u, v)$ by, 
\[\begin{cases}
u\equiv& t-r_\ast=\text{const.} \\
&\text{(represents "outgoing' null geodesics. or phase of the outgoing mode)}\\
v\equiv& t+r_\ast=\text{const.} \\
&\text{(represents "incoming" null geodesic. or phase of the incoming mode)}
\end{cases}\]
First, using
\[
\begin{cases}
du=dt-dr_\ast=dt-\left(\frac{dr_\ast}{dr}\right)dr=dt-e^{\left(\frac{c^2}{2r^2}\right)}\left[\frac{r}{r^2-2GMr+(4\pi GQ^2+c^2)}\right]dr\\
dv=dt+dr_\ast=dt+\left(\frac{dr_\ast}{dr}\right)dr=dt+e^{\left(\frac{c^2}{2r^2}\right)}\left[\frac{r}{r^2-2GMr+(4\pi GQ^2+c^2)}\right]dr
\end{cases}
\]
$\therefore dudv=dt^2-\frac{1}{e^{\left(-\frac{c^2}{r^2}\right)}\left[1-\frac{2GM}{r}+\frac{(4\pi GQ^2+c^2)}{r^2}\right]^2}dr^2$\\\\
and hence in terms of these null coordinates,
\[ds^2=-e^{\left(-\frac{c^2}{r^2}\right)}\left[1-\frac{2GM}{r}+(4\pi GQ^2+c^2)\frac{1}{r^2}\right]dudv\]
Further, using
\begin{align*}
\left(\frac{v-u}{2}\right)&=r_\ast=e^{\left(\frac{c^2}{2r^2}\right)}\left\{r+\left(\frac{r^2_+}{r_+-r_-}\right)\ln(r-r_+)(r-r_-)-\left(\frac{r^2_+-r^2_-}{r_+-r_-}\right)\ln(r-r_-)\right\}\\
e^{\left(-\frac{c^2}{2r^2}\right)}\frac{v-u}{\left(\frac{2r^2_+}{r_+-r_-}\right)}&=\frac{r}{\left(\frac{r^2_+}{r_+-r_-}\right)}+\ln(r-r_+)(r-r_-)-\left(\frac{r^2_+-r^2_-}{r^2_+}\right)\ln(r-r_-)\\
\exp\left[e^{\left(-\frac{c^2}{2r^2}\right)}\frac{(v-u)}{\left(\frac{2r^2_+}{r_+-r_-}\right)}\right]&=e^{\left[r\big/\left(\frac{r^2_+}{r_+-r_-}\right)\right]}e^{-\left(\frac{r^2_+-r^2_-}{r^2_+}\right)\ln(r-r_-)}(r-r_+)(r-r_-)\\
&=e^{\left[r\big/\left(\frac{r^2_+}{r_+-r_-}\right)\right]}(r-r_-)^{-\left(\frac{r^2_+-r^2_-}{r^2_+}\right)}\left[\frac{r^2-2GMr+(4\pi GQ^2+c^2)}{r^2}\right]r^2.
\end{align*}
Thus, the metric can be further rewritten as,
\[ds^2=-e^{\left(-\frac{c^2}{r^2}\right)}\frac{1}{r^2}(r-r_-)^{\left(\frac{r^2_+-r^2_-}{r^2_+}\right)}e^{-\left[r\big/\left(\frac{r^2_+}{r_+-r_-}\right)\right]}\exp\left[e^{\left(-\frac{c^2}{2r^2}\right)}\right]\frac{(v-u)}{\left(\frac{2r^2_+}{r_+-r_-}\right)}dudv.  \]
Now, go to "another choice of null coordinates", $(U,\ V)$ defined by,
\[
\begin{cases}
U\equiv-\exp\left[-e^{\left(-\frac{c^2}{2r^2}\right)}\frac{u}{\left(\frac{2r^2_+}{r_+-r_-}\right)}\right]\\
V\equiv\exp\left[e^{\left(-\frac{c^2}{2r^2}\right)}\frac{v}{\left(\frac{2r^2_+}{r_+-r_-}\right)}\right]
\end{cases}
\]
\[\Rightarrow
\begin{cases}
dU=\left(\frac{r_+-r_-}{2r^2_+}\right)\exp\left[-e^{\left(-\frac{c^2}{2r^2}\right)}\frac{u}{\left(\frac{2r^2_+}{r_+-r_-}\right)}\right]d\left(e^{\left(-\frac{c^2}{2r^2}\right)}u\right)\\
dV=\left(\frac{r_+-r_-}{2r^2_+}\right)\exp\left[e^{\left(-\frac{c^2}{2r^2}\right)}\frac{v}{\left(\frac{2r^2_+}{r_+-r_-}\right)}\right]d\left(e^{\left(-\frac{c^2}{2r^2}\right)}v\right)\
\end{cases}
\]
Here, note that
\begin{align*}
&d\left(e^{\left(-\frac{c^2}{2r^2}\right)}u\right)d\left(e^{\left(-\frac{c^2}{2r^2}\right)}v\right)\\
&=\left[e^{\left(-\frac{c^2}{2r^2}\right)}du+\frac{c^2}{r^3}ue^{\left(-\frac{c^2}{2r^2}\right)}dr\right]\left[e^{\left(-\frac{c^2}{2r^2}\right)}dv+\frac{c^2}{r^3}ve^{\left(-\frac{c^2}{2r^2}\right)}dr\right]\\
&=e^{\left(-\frac{c^2}{r^2}\right)}\left[dudv+\frac{c^2}{r^3}(vdu+udv)dr+\frac{c^4}{r^6}(uv)dr^2\right].\\
\end{align*}
\begin{align*}
\therefore\ e^{\left(-\frac{c^2}{r^2}\right)}dudv=&d\left(e^{\left(-\frac{c^2}{2r^2}\right)}u\right)d\left(e^{\left(-\frac{c^2}{2r^2}\right)}v\right)-e^{\left(-\frac{c^2}{r^2}\right)}\left[\frac{c^2}{r^3}(vdu+udv)dr+\frac{c^4}{r^6}(uv)dr^2\right]\\
=&d\left(e^{\left(-\frac{c^2}{2r^2}\right)}u\right)d\left(e^{\left(-\frac{c^2}{2r^2}\right)}v\right)+\mathcal{O}\left(\frac{1}{r^2}\right)dr^2.
\end{align*}
where we used: \\\\
$\begin{cases}
du\sim dv\sim e^{\left(\frac{c^2}{2r^2}\right)}dr\\
u\sim v\sim r_\ast\sim re^{\left(\frac{c^2}{2r^2}\right)}\\
(vdu)\sim(udv)\sim re^{\left(\frac{c^2}{r^2}\right)}dr\\
(uv)\sim r^2e^{\left(\frac{c^2}{r^2}\right)}
\end{cases}$\\\\
Thus, we have,
\begin{align*}
&\exp\left[e^{\left(-\frac{c^2}{2r^2}\right)}\frac{(v-u)}{\left(\frac{2r^2_+}{r_+-r_-}\right)}\right]e^{\left(-\frac{c^2}{r^2}\right)}dudv\\
&=\left(\frac{2r^2_+}{r_+-r_-}\right)^2dUdV+\exp\left[e^{\left(-\frac{c^2}{2r^2}\right)}\frac{(v-u)}{\left(\frac{2r^2_+}{r_+-r_-}\right)}\right]\mathcal{O}\left(\frac{1}{r^2}\right)dr^2.
\end{align*}
Therefore, the metric can be rewritten again as,
\[ds^2=-\left(\frac{2r^2_+}{r_+-r_-}\right)^2\frac{1}{r^2}(r-r_-)^{\left(\frac{r^2_+-r^2_-}{r^2_+}\right)}e^{-\left[r\big/\left(\frac{r^2_+}{r_+-r_-}\right)\right]}dUdV+\mathcal{O}\left(\frac{1}{r^{3-4}}\right)dr^2.\]
Notice, here, that dropping the terms of order $\mathcal{O}\left(\frac{1}{r^{3-4}}\right)$ is indeed consistent with our earlier approximation in which the terms $\mathcal{O}\left(\frac{1}{r^2}\right)$ have been neglected in the metric written in terms of original coords. $(t, r).$\\\\
Finally, we define (generalized) "Kruskal coordinates" $(T, X)$ by,\\

$\begin{cases}
T\equiv\frac{1}{2}(V+U)=\exp\left[e^{\left(-\frac{c^2}{2r^2}\right)}\left(\frac{r_+-r_-}{2r^2_+}\right)r_\ast\right]\sinh\left[e^{\left(-\frac{c^2}{2r^2}\right)}\left(\frac{r_+-r_-}{2r^2_+}\right)t\right]\\
X\equiv\frac{1}{2}(V-U)=\exp\left[e^{\left(-\frac{c^2}{2r^2}\right)}\left(\frac{r_+-r_-}{2r^2_+}\right)r_\ast\right]\cosh\left[e^{\left(-\frac{c^2}{2r^2}\right)}\left(\frac{r_+-r_-}{2r^2_+}\right)t\right]
\end{cases}$\\
$\Rightarrow$$\quad\begin{cases} U=T-X\\V=T+X \end{cases}$ $\Rightarrow$\quad $dUdV=dT^2-dX^2$.\\\\
Finally, in terms of these (generalized) "Kruskal coords" (T, X, $\theta$, $\phi$), the full Reissner-Nordstr{\"o}m metric is written as:\\
\[ds^2=\left(\frac{2r^2_+}{r_+-r_-}\right)^2\frac{1}{r^2}(r-r_-)^{\left(\frac{r^2_+-r^2_-}{r^2_+}\right)}e^{-\left[r\big/\left(\frac{r^2_+}{r_+-r_-}\right)\right]}(-dT^2+dX^2)+r^2d\Omega^2_2\].\\
\textbf{Relations between the old coords. $(t, r)$ and the new, Kruskal coords. $(T, X)$}\\
Consider, 
\begin{align*}
X^2-T^2&=\exp\left[e^{\left(-\frac{c^2}{2r^2}\right)}\left(\frac{r_+-r_-}{r^2_+}\right)r_\ast\right]\\
&=\exp\left[r\bigg/\left(\frac{r^2_+}{r_+-r_-}\right)+\ln(r-r_+)(r-r_-)-\left(\frac{r^2_+-r^2_-}{r^2_+}\right)\ln(r-r_-)\right]
\end{align*}
thus,
\[X^2-T^2=e^{\left[r\big/\left(\frac{r^2_+}{r_+-r_-}\right)\right]}(r-r_-)^{-\left(\frac{r^2_+-r^2_-}{r^2_+}\right)}(r-r_+)(r-r_-)\cdots\cdots\textcircled{A}\]\\
and consider,
\begin{align*}
t=\frac{1}{2}(u+v)&=e^{\left(\frac{c^2}{2r^2}\right)}\left(\frac{r^2_+}{r_+-r_-}\right)\left[-\ln(-U)+\ln V\right]\\
&=e^{\left(\frac{c^2}{2r^2}\right)}\left(\frac{r^2_+}{r_+-r_-}\right)\ln\left(\frac{V}{-U}\right)=e^{\left(\frac{c^2}{2r^2}\right)}\left(\frac{r^2_+}{r_+-r_-}\right)\ln\left(\frac{X+T}{X-T}\right)\\
&=e^{\left(\frac{c^2}{2r^2}\right)}\left(\frac{2r^2_+}{r_+-r_-}\right)\tanh^{-1}\left(\frac{T}{X}\right)
\end{align*}
thus, 
\[\frac{T}{X}=\tanh\left[e^{\left(-\frac{c^2}{2r^2}\right)}\left(\frac{r_+-r_-}{2r^2_+}\right)t\right]\cdots\cdots\textcircled{B}\]
(where we used $\tanh^{-1}(z)=\frac{1}{i}\tan^{-1}(iz)=\frac{1}{2}\ln\left(\frac{1+z}{1-z}\right))$\\

\noindent\textbf{[Note]} On the structure of event horizons\\
Concerning the event horizons, their structure may be of particular interest.\\
Just as the Reissner-Nordstr{\"o}m metric solution in general relativity, our monopole black hole metric has both inner and outer horizons. However, due to the presence of the dilaton field, these two horizons of our monopole black hole get closer to each other than the usual Reissner-Nordstr{\"o}m black hole horizons (namely, $r^{ours}_->r^{RN}_-,\  r^{ours}_+<r^{RN}_+$). And the spacetime geometry is not singular at either of these two event horizons as is obvious in the form of our monopole black hole metric expressed in terms of Kruskal coordinates.\\
    Next, it may be worth noting that the inner horizon is unstable whereas the outer horizon is stable in that non-spherically symmetric (i.e. anisotropic) perturbations tend to blow up on the inner event horizon as is well-known in the case of the usual Reissner-Nordstr{\"o}m $\text{metric}^\ast.$\\
($\ast$ S. Chandrasekhar and J. Hartle, Proc. R. Soc. London A384, 301 (1982), and references therein.)\\
Finally, notice that the physical distance to the event horizon is finite although it can be traversed in finite proper time for our monopole black hole, namely
\[
L(r_H)=\int^{r_H}_0\sqrt{g_{rr}}dr
\]
diverges since $\sqrt{g_{rr}}$ behaves as $r(r-r_H)^{-1}$ near the horizon. This fact, then, implies that all the non-trivial field configurations representing the 'entire' structure of our non-Abelian monopole take place only within the event horizon in accordance with the "no-hair" theorems!\\\\\
\noindent\textbf{[Note]} The relationship $\textcircled{A}$ shows that the spatial coordinate $r$ actually is an implicit function of $T$ which is a "global time" coordinate. Therefore, in fact the (dilatonic gravity's version of) Reissner-Nordstr{\"o}m metric does depend on time and hence the spacetime is not static globally leading to the "Black hole evaporation" via the emission of (quantum) particle $\cdots\cdots$ Hawking Radiation!\\
Now in the same manner as one explores the Hawking radiation of Schwarzschild black hole, one can readily read off the "Hawking temperature", $T_H$ from the periodicity in time coordinate of the Euclidean section. (viz.)\\
By "analytic continuation" $(\tau=it)$, we go to the Euclidean signature in which,\\
\begin{align*}
T&=\exp\left[e^{\left(-\frac{c^2}{2r^2}\right)}\left(\frac{r_+-r_-}{2r^2_+}\right)r_\ast\right]\sin\left[e^{\left(-\frac{c^2}{2r^2}\right)}\left(\frac{r_+-r_-}{2r^2_+}\right)\tau\right](-i)\\
&\equiv(-i)T_E\\
X&=\exp\left[e^{\left(-\frac{c^2}{2r^2}\right)}\left(\frac{r_+-r_-}{2r^2_+}\right)r_\ast\right]\cos\left[e^{\left(-\frac{c^2}{2r^2}\right)}\left(\frac{r_+-r_-}{2r^2_+}\right)\tau\right]\\
&\equiv X_E
\end{align*}
where we used\\ 
$\sinh(t)=\sinh(-i\tau)=(-i)\sin(\tau),\\
\cosh(t)=\cosh(-i\tau)=\cos(\tau)$\\\\
Thus the "Wick rotation" of old coordinate, $\tau=it$ is equivalent to the Wick rotation of new, Kruskal-like coordinate, $T_E=iT$ (and of course, $dT_E=idT$).\\
Then the "positive-definite" Euclidean metric reads,
\[
ds^2_E=\left(\frac{2r^2_+}{r_+-r_-}\right)\frac{1}{r^2}(r-r_-)^{\left(\frac{r^2_+-r^2_-}{r^2_+}\right)}e^{-\left[ r\big/\left(\frac{r^2_+}{r_+-r_-}  \right) \right]}(dT^2_E+dX^2)+r^2d\Omega^2_2.
\]
Now, obviously, the Einstein time coordinate, $\tau$ has a periodicity of $T=\frac{2\pi}{\omega}\simeq2\pi e^{\left(\frac{c^2}{2r^2}\right)}\left(\frac{2r^2_+}{r_+-r_-}\right).$ Thus by identifying $\beta=T=\frac{1}{T_H}$, we obtain the "Hawking temperature" to be.
\begin{align*}
T_H&\simeq\frac{1}{2\pi}e^{\left(-\frac{c^2}{2r^2}\right)}\left(\frac{r_+-r_-}{2r^2_+}\right)\\
&=\frac{1}{2\pi G}\frac{\sqrt{M^2-\frac{1}{G^2}(4\pi GQ^2+c^2)}}{\left[M+\sqrt{M^2-\frac{1}{G^2}(4\pi GQ^2+c^2)}\right]^2}e^{\left(-\frac{c^2}{2r^2} \right)}
\end{align*}
First, note that even to the leading order in our approximation in solving the Einstein equations, the Hawking temperature has a dependence on the $r$-coordinate in such a manner that $T_H$ gets higher as $r$ gets larger for fixed $M$ implying that the Hawking temp. near the surface of our monopole black hole is higher than that inside of it! (But, of course, this analysis is valid only in the "large-$r$" region.)

\section{Proof of the inequalities in Eq.(5.4)}
\label{inequ}

\noindent(i) First inequality\\
: We begin by solving the second combination of Einstein equations.\\
$\Rightarrow$ Eq.(5.2) can be recast into the form\\
\[
\begin{split}
&M'(r)+8\pi GrK(r)M(r)=4\pi r^2[K(r)+V(r)]\\
&\frac{d}{dr}[M(r)e^{-I(r)}]=4\pi r^2[K(r)+V(r)]e^{-I(r)}
\end{split}
\]
where\quad $I(r)\equiv\int^\infty_rdr'8\pi Gr'K(r')$\\
which, upon integration, yields
\[
\begin{split}
&M(r)e^{-I(r)}-M(0)e^{-I(0)}=4\pi\int^r_0dr r^2[K(r)+V(r)]e^{-I(r)}\\
&\therefore M(r)=\left\{4\pi\int^r_0dr'r'^2[K(r')+V(r')]e^{-I(r')}+M(0)e^{-I(0)}\right\}e^{I(r)}
\end{split}
\]
Thus the total mass (energy) of the monopole is,\\
\[M\equiv M(\infty)=4\pi\int^\infty_0dr\ r^2[K(r)+V(r)]e^{-I(r)}+M(0)e^{-I(0)}\]
Now, since $K(r)$ is positive-definite,\\
\[I(r)\le I(0)\]
and hence
\[M=M(\infty)\ge\left\{4\pi\int^\infty_0dr\ r^2[K(r)+V(r)]+M(0)\right\}e^{-I(0)}\]
where the equality holds if $K(r)=0$, namely if the 3 positive-definite terms in $K(w, u, h)$ given in Eq.(3.8) vanish separately,
\[
\begin{cases}
w'(r)+\frac{1}{r}w(r)=0\\
u'(r)=0\\
h'(r)=0
\end{cases}
\]\\
(ii) Second inequality
Note first that (in the sign convention ($-+++$)):
\begin{align*}
\frac{1}{16\pi G}\left\{-2(\nabla\Phi)^2\right\}&=-\frac{2}{16\pi G}\left\{g^{\mu\nu}(D_\mu\Phi)(D_\nu\Phi)\right\}\\
&=-\frac{2}{16\pi G}\left\{g^{00}(D_0\Phi)^2+g^{ii}(D_i\Phi)^2\right\}\\
&=-\frac{2}{16\pi G}\left\{g^{rr}(D_r\Phi)^2+g^{00}(D_0\Phi)^2+g^{\phi\phi}(D_\phi\Phi)^2\right\}\\
&=-\frac{1}{8\pi GM^2}\left\{\left(\frac{1}{A}\right)\left(\frac{w'}{r}-\frac{w}{r^2}\right)^2\right\}\\
-\frac{1}{4}(F^a_{\mu\nu})^2&=-\frac{1}{4}g^{\mu\alpha}g^{\nu\beta}F^a_{\mu\nu}F^a_{\alpha\beta}=-\frac{1}{4}\left\{2g^{00}g^{ii}F^a_{0i}F^a_{0i}+2g^{ii}g^{jj}F^a_{ij}F^a_{ij}\right\}\\
&=-\frac{1}{2}\left\{g^{00}g^{ii}(E^a_i)^2+g^{ii}g^{ij}(\epsilon_{ijk}B^a_k)^2\right\}\\
&=-\frac{1}{2}\left\{g^{\theta\theta}g^{\phi\phi}(B^a_r)^2+g^{\phi\phi}g^{rr}(B_\theta)^2+g^{rr}g^{\theta\theta}(B_\phi)^2\right\}\\
&=-\left\{\left(\frac{1}{A}\right)\frac{(u')^2}{e^2r^2}+\frac{(u^2-1)^2}{2e^2r^4}\right\}\\
-\frac{1}{2}e^{2\Phi}(D_\mu\phi^a)^2&=-\frac{1}{2}e^{2\Phi}\left\{g^{\mu\nu}(D_\mu\phi^a)(D_\nu\phi^a)\right\}\\
&=-\frac{1}{2}e^{2\Phi}\left\{g^{00}(D_0\phi^a)^2+g^{ii}(D_i\phi^a)^2\right\}\\
&=-\frac{1}{2}e^{2\Phi}\left\{g^{rr}(D_r\phi^a)^2+g^{\theta\theta}(D_\theta\phi^a)^2+g^{\phi\phi}(D_\phi\phi^a)^2\right\}\\
&=-\frac{1}{2}e^{2\left(\frac{1}{Mr}w+\Phi_\infty\right)}\left\{\left(\frac{1}{A}\right)\frac{v^2(h')^2}{2}+\frac{v^2h^2u^2}{r^2}\right\}\\
e^{4\Phi}U(\phi^a\phi^a)&=e^{4\left(\frac{1}{Mr}w+\Phi_\infty\right)}\frac{\lambda}{4}v^4(h^2-1)^2.
\end{align*}
And
\begin{align*}
&K(w,u,h)=\frac{1}{8\pi GM^2}\left(\frac{w'}{r}-\frac{w}{r^2}\right)^2+\frac{(u')^2}{e^2r^2}+e^{2\left(\frac{1}{Mr}w+\Phi_\infty\right)}\frac{v^2(h')^2}{2}\\
&V(w,u,h)=\frac{(u^2-1)^2}{2e^2r^r}+e^{2\left(\frac{1}{Mr}w+\Phi_\infty\right)}\frac{v^2h^2u^2}{r^2}+e^{4\left(\frac{1}{Mr}w+\Phi_\infty\right)}\frac{\lambda}{4}v^4(h^2-1)^2
\end{align*}
$\Rightarrow$ Thus,
\begin{align*}
\rho_m(r)&=(-T^0_0)\\
&=\left[\frac{2}{16\pi G}g^{ii}(D_i\Phi)^2+\frac{1}{2}g^{ii}g^{jj}(\epsilon_{ijk}B^a_k)^2+\frac{1}{2}e^{2\Phi}g^{ii}(D_i\phi^a)^2+e^{4\Phi}U(\phi^a\phi^a)\right]\\
&=\left[\left(\frac{1}{A}\right)K+V\right].\\
\end{align*}
Notice that the difference between the monopole mass (energy) density in flat and in curved spacetime arises due to the non-trivial metric
\[
g_{rr}=\begin{cases}
1\quad (\text{flat})\\
A(r)\quad (\text{curved})
\end{cases}
\]
Namely, in flat spacetime,
\begin{align*}
\rho_m(r)&=\left(-T^0_0\right)\\
&=\left[\frac{2}{16\pi G}(D_i\Phi)^2+\frac{1}{2}(B^a_i)^2+\frac{1}{2}e^{2\Phi}(D_i\phi^a)^2+e^{4\Phi}U(\phi^a\phi^a)\right]\\
&=\left[K+V\right]
\end{align*}
$\Rightarrow$ Thus if $M(0)=0$ as is needed for the monopole to be non-singular,
\begin{align*}
M&=M(\infty)\ge\left\{4\pi\int^\infty_0dr\ r^2[K(r)+V(r)]\right\}e^{-I(0)}\\
&=\left\{\int d^3x\left[\frac{2}{16\pi G}(D_i\Phi)^2+\frac{1}{2}(B^a_i)^2+\frac{1}{2}e^{2\Phi}(D_i\phi^a)^2+e^{4\Phi}U(\phi^a\phi^a)\right]\right\}e^{-I(0)}\\
&\ge\left\{\int d^3x\left[\frac{1}{2}(B^a_i)^2+\frac{1}{2}(D_i\phi^a)^2+U(\phi^a\phi^a)\right]\right\}e^{-I(0)}\\
&\ge\left\{\int d^3x\left[\frac{1}{2}(B^a_i)^2+\frac{1}{2}(D_i\phi^a)^2\right]\right\}e^{-I(0)}\\
&\ge\left\{\int d^3x\left[B^a_i-(D_i\phi^a) \right]\right\}e^{-I(0)}\\
&=\left\{\int d^3x\left[D_i(B^a_i\phi^a)-(D_iB^a_i)\phi^a\right]\right\}e^{-I(0)}\\
&=\left\{\int ds^i(B^a_i\phi^a)\right\}e^{-I(0)}=\left\{v\int ds^i(B^a_i\hat{\phi}^a)\right\}e^{-I(0)}\\
&=(vQ_{\text{mag}})e^{-I(0)}=\left(\frac{4\pi v}{e}\right)e^{-I(0)}.
\end{align*}
Namely the second inequality holds if the following Bogomol'nyi equation is satisfied: $\Phi=0,\ U(\phi^a\phi^a)=0,\ B^a_i=\pm(D_i\phi^a)$\\
In terms of the spherically symmetric ansatz, this Bogomol'nyi condition can be written as,
\begin{align*}
w(r)&=-Mr(\Phi_\infty),\ \lambda=0\\
(u^2-1)&=-evh'(r)\\
u'(r)&=-evh(r)u(r)
\end{align*}
Consequently, in principle the solution to this Bogomol'nyi equations Eq.(5.6) and Eq.(5.7) would be the minimum energy configuration that saturates the Bogomol'nyi bound. As P. van Nieuwenhuizen et al. put it, that our curved spacetime monopole solution has a positive-definite lower bound on its total mass physically means that although bringing matter closely together yields a large negative Newtonian energy, this is compensated for by the positive energy which resides in the curving of space.


\begin{thebibliography}{99}
\bibitem{1} F.A. Bais and R. J. Russell, Phys. Rev. D13, 2692 (1975); Y.M. Cho and P.G.O. Freund, ibid, 12, 1588 (1975).
\bibitem{2} P. van Nieuwenhuizen, D. Wilkinson and M.J. Perry, phys. Rev. D13, 778 (1976).
\bibitem{3} K. Lee, V.P. Nair and E.J. Weinberg, Phys. Rev. D45, 2751 (1992);\\Phys. Rev. Lett. 68, 1100 (1992); M.E. Ortiz, Phys. Rev. D45, 2585 (1992).
\bibitem{4} G. 't Hooft, Nucl. Phys. B79, 276 (1974);\\
A.M. Polyakov, Pis'ma Zh. Eksp. Teor. Fiz.20, 430 (1974) [JETP Lett. 20,194 (1974)];\\
P. Goddard and D. Olive, Rep. Prog. Phys. 41, 1357 (1978).
\bibitem{5} G.W. Gibbons, Self-gravitating magnetic monopoles, global monopoles and black holes, Lectures 1990 Lisbon Autumn school. and References therein; G.W. Gibbons and K. Maeda, Nucl. Phys. B298,741 (1998); D. Garfinkle, G. Horowitz and A. Strominger, Phys. Rev. D43, 3140 (1991).
\bibitem{6} A. Strominger, Nucl. Phys. B343, 167 (1990); B353, 565(E) (1991);\\
S.J. Rey, Phys. Rev. D43, 439 (1991); T. Banks, M. Dine, H. Dijkstra and W. Fischler, Phys. Lett. B212, 45 (1998).
\bibitem{7} J.A. Harvey and J. Lin, Phys. Lett. B268, 40 (1991).
\bibitem{8} M. Dine and N. Seiberg, Phys. Rev. Lett. 55, 366 (1985); C.G. Callan, D. Friedan, D.J. Martines and M.J. Perry, Nucl. Phys. B262, 593 (1985); G.W. Gibbons and K. Maeda, Nucl. Phys. B298, 741 (1998).
\bibitem{9} H. Georgi and S.L. Glashow, Phys. Rev. Lett. 32, 438 (1974).
\bibitem{10} E. Bogomol'nyi, Yad. Fiz. 24, 861 (1975) [Sov. J. Nucl. Phys. 24, 449 (1975)]; S. Coleman, S. Parke. A. Neveu and C. Sommerfield, Phys. Rev. D15, 544 (1977).
\bibitem{11} S. Chandrasekhar and J. Hartle, Proc. R. Soc. London A384, 301 (1982) and references therein.
\bibitem{12} J. D. Bekenstein, Phys. Rev. D5, 1239 (1972);\\
S.W. Hawking, Comm. Math. Phys. 25, 152 (1972); W. Israel, ibid. D12, 245 (1968).
\bibitem{13} S.W. Hawking, Comm. Math. Phys. 43, 199 (1975); L. Parker, Phys. Rev. D12, 1519 (1975).
\bibitem{14} B. carter, in Black Holes, eds. C. Dewitt and B.S. Dewitt (New York, 1975, Gordon \& Breach); N.D. Birrell and P.C.W. Davis, Quantum Fields in Curved space (Cambridge University Press, Cambridge, England, 1982).
\end{thebibliography}
\end{document}